\documentclass[iop,twocolappendix]{emulateapj}
\usepackage{apjfonts}
\usepackage{amsmath,amssymb}
\usepackage{epstopdf}

\renewcommand{\vec}[1]{\mathbf{#1}}
\newcommand{\diffp}[2]{\frac{\partial #1}{\partial #2}}
\newcommand{\diff}[2]{\frac{d #1}{d #2}}

\shorttitle{Collisionless Isotropization of the Solar-Wind Protons}
\shortauthors{D.~Verscharen et al.}

\begin{document}

\title{Collisionless Isotropization of the Solar-Wind Protons by Compressive Fluctuations\\ and Plasma Instabilities}


\author{Daniel Verscharen$^{1}$, Benjamin D.~G.~Chandran$^{1,3}$, Kristopher G.~Klein$^{1,4}$, and Eliot Quataert$^{2}$}
\affil{\mbox{$^{1}$Space Science Center, University of New Hampshire, Durham, NH 03824, USA; daniel.verscharen@unh.edu, benjamin.chandran@unh.edu, kriskl@umich.edu}\\
\mbox{$^{2}$Astronomy Department and Theoretical Astrophysics Center, 501 Campbell Hall, The University of California, Berkeley, CA 94720, USA; eliot@astro.berkeley.edu}}
\altaffiltext{3}{Also at Department of Physics, University of New Hampshire, Durham, NH 03824, USA}
\altaffiltext{4}{Now at Department of Atmospheric, Oceanic, and Space Sciences, University of Michigan, Ann Arbor, MI 48109, USA}

\journalinfo{The Astrophysical Journal, 831:128 (11pp), 2016 November 10}
\submitted{Received 2016 May 23; revised 2016 August 9; accepted 2016 August 10; published 2016 November 3}

\begin{abstract}
Compressive fluctuations are a minor yet significant component of astrophysical plasma turbulence. In the solar wind, long-wavelength compressive slow-mode fluctuations lead to changes in $\beta_{\parallel \mathrm p}\equiv 8\pi n_{\mathrm p}k_{\mathrm B}T_{\parallel \mathrm p}/B^2$ and in $R_{\mathrm p}\equiv T_{\perp \mathrm p}/T_{\parallel \mathrm p}$, where $T_{\perp \mathrm p}$ and $T_{\parallel \mathrm p}$ are the perpendicular and parallel temperatures of the protons, $B$ is the magnetic field strength, and $n_{\mathrm p}$ is the proton density. 
If the amplitude of the compressive fluctuations is large enough, $R_{\mathrm p}$  crosses one or more instability thresholds for anisotropy-driven microinstabilities.
The enhanced field fluctuations from these microinstabilities scatter the protons so as to reduce the anisotropy of the pressure tensor. We propose that this scattering drives the average value of $R_{\mathrm p}$ away from the marginal stability boundary until the fluctuating value of $R_{\mathrm p}$ stops crossing the boundary.  We model this ``fluctuating-anisotropy effect'' using linear Vlasov--Maxwell theory to describe the large-scale compressive fluctuations. We argue that this effect can explain why, in the nearly collisionless solar wind, the average value of $R_{\mathrm p}$ is  close to unity. 
\end{abstract}


\section{Introduction}

Astrophysical plasmas are often in a turbulent state \citep{alexandrova13,bruno13}. This turbulence is characterized by a broad distribution of fluctuations over a wide range of wavevectors and by an ongoing turbulent cascade that transfers energy between different wavevectors \citep{schekochihin09,wicks11,carbone12}. A background magnetic field $\vec B_0$ can lead to an anisotropic cascade \citep{goldreich05}, and, in fact, most astrophysical plasmas are permeated by such a magnetic field. Theoretical considerations, numerical simulations, and solar-wind observations indicate that the turbulent cascade -- especially for large-scale noncompressive fluctuations -- preferentially transfers energy to large $k_{\perp}$, and to a lesser extent to large $k_{\parallel}$ in a magnetized plasma, where $k_{\perp}$ ($k_{\parallel}$) is the wavenumber in the direction perpendicular (parallel) to $\vec B_0$ \citep{montgomery81,oughton94,oughton98,cho00,sahraoui10,chen11, narita11,he12,salem12,verscharen12}. 
We focus our treatment on the outer scales of the turbulence at which the frequencies (linear and nonlinear) are $\ll \Omega_{\mathrm p}$, where $\Omega_{\mathrm p}\equiv q_{\mathrm p}B_0/m_{\mathrm p}c$ is the proton gyrofrequency, $q_{\mathrm p}$ and $m_{\mathrm p}$ are the charge and the mass of a proton, and $c$ is the speed of light.

In addition to the anisotropic cascade,  properties of the particle distribution function can be anisotropic with respect to $\vec B_0$ in a plasma with low collisionality. For example, in situ measurements of the solar wind have shown temperature anisotropies with $R_{\mathrm p}\equiv T_{\perp \mathrm p}/T_{\parallel \mathrm p}\neq 1$, where $T_{\perp \mathrm p}$ ($T_{\parallel \mathrm p}$) is the temperature of the protons in the direction perpendicular (parallel) to the magnetic field \citep{marsch82,kasper02,hellinger06,marsch06,bale09,maruca12}.
If the temperature anisotropy $|R_{\mathrm p}-1|$ of the protons exceeds a certain threshold, then the plasma becomes unstable, and various kinds of plasma waves and/or nonpropagating structures grow, while the distribution function relaxes toward a stable state. If $R_{\mathrm p}>1$, the plasma can excite parallel-propagating Alfv\'en/ion-cyclotron (A/IC) waves or nonpropagating mirror modes \citep{rudakov61,sagdeev61,tajiri67,southwood93,gary94,kunz14,riquelme15,gary16}. If $R_{\mathrm p}<1$, the plasma can excite parallel-propagating fast-magnetosonic/whistler (FM/W) waves or nonpropagating oblique firehose modes \citep{quest96,gary98,hellinger00,hellinger08,rosin11}. The anisotropy thresholds of these instabilities decrease with increasing $\beta_{\parallel \mathrm p}\equiv 8\pi n_{\mathrm p} k_{\mathrm B}T_{\parallel \mathrm p}/B^2$, where $k_{\mathrm B}$ is the Boltzmann constant, $n_{\mathrm p}$ is the proton density, and $\vec B$ is the magnetic field. According to the double-adiabatic or Chew--Goldberger--Low \citep[CGL;][]{chew56} model, the solar wind is expected to develop strong temperature anisotropy with $R_{\mathrm p}<1$ during its transit from the Sun to a heliocentric distance $r$ of 1 au, so that the plasma would approach the marginal-stability curves of the anisotropy instabilities at $R_{\mathrm p}<1$ in the $\beta_{\parallel\mathrm p}$-$R_{\mathrm p}$ plane. Some models \citep[e.g.,][]{hellinger08,chandran11,yoon14} have suggested that, after the solar wind first encounters the marginal-stability curve, it evolves along the marginal-stability curve as it moves away from the Sun since anisotropy instabilities prevent the plasma from moving past the marginal-stability curve. However, numerous observations show that the solar wind exhibits a broad distribution of $R_{\mathrm p}$ values that is peaked at $R_{\mathrm p}\simeq 1$, even in wind streams with very low collisionality \citep[e.g.,][]{marsch82,bale09}. This implies that an additional physical mechanism counteracts the double-adiabatic reduction in $R_{ \mathrm p}$. One potential explanation for this puzzle is the existence of a perpendicular heating mechanism with just the right magnitude to offset the tendency of adiabatic expansion to decrease $R_{\mathrm p}$. We explore an alternative ansatz by investigating the possibility that compressive fluctuations lead to a reduction of the anisotropy away from the instability thresholds.

In this paper, we propose the following basic concept as a mechanism for the collisionless isotropization of the plasma. The expansion of the solar wind in the inner heliosphere ($r\lesssim 1\,\mathrm{au}$) leads to a reduction in the average value of $R_{\mathrm p}$, which we denote $R_{0\mathrm p}$, and an increase in the equilibrium value of $\beta_{\parallel\mathrm p}$, which we denote  $\beta_{\parallel 0\mathrm p}$. The presence of large-scale slow-mode fluctuations with amplitude $\delta |\vec B|$ leads to fluctuations in $R_{\mathrm p}$ around $R_{0\mathrm p}$ as well as fluctuations in $\beta_{\parallel\mathrm p}$ around $\beta_{\parallel 0\mathrm p}$. At some distance from the Sun, $R_{0\mathrm p}$ sufficiently decreases that the fluctuating value of $R_{\mathrm p}$ crosses the threshold for the FM/W instability. The instability then increases $R_{0\mathrm p}$ via  pitch-angle scattering until the fluctuating value of $R_{\mathrm p}$ stops crossing the instability threshold. As the plasma expands further, this ``fluctuating-anisotropy effect'' keeps $R_{0\mathrm p}$ at a sizable ``distance'' from the marginal-stability boundary, in agreement with observations that show that $R_{0\mathrm p}$ is close to unity even when collisions are weak. Our work aims to introduce this novel physical mechanism and is largely based on conceptual arguments. In order to test rigorously whether this mechanism is important in the solar wind, additional numerical and observational studies will be necessary.

The goal of our present work is to develop a quantitative description of the limits on $\beta_{\parallel \mathrm p}$ and $R_{\mathrm p}$  that are set by the FM/W instability and large-scale compressions with different amplitudes $\delta |\vec B|/B_0$. We note that in other circumstances (e.g., accretion flows in which double-adiabatic compression acts to increase $R_{0\mathrm p}$), this mechanism could place an upper limit on $R_{0\mathrm p}$ through the combined action of compressions and the A/IC instability. 

Previous treatments based their analysis on the notion that microscale plasma instabilities triggered by time-dependent fluctuations in the magnetic field strength pin the temperature anisotropy at its marginally stable value \citep[e.g., ][]{schekochihin05,schekochihin06,sharma06,kunz11}. In contrast, we focus on the possibility that transient encounters between the time-evolving value of $R_{\mathrm p}$ and an instability threshold drive $R_{0\mathrm p}$ toward isotropy.

 \begin{figure}
\epsscale{1.2}
\plotone{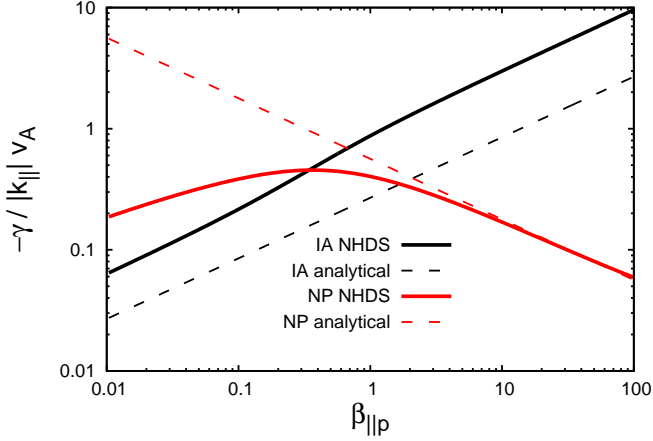}
\caption{Damping rates of the IA mode and of the NP mode in numerical solutions obtained with NHDS and from our analytical theory according to Equations~(\ref{gammaIAW}) and (\ref{gammaNPM}). We use the parameters $T_{\parallel\mathrm p}=T_{\parallel\mathrm e}$, $\theta=88^{\circ}$, $k_{\parallel}v_{\mathrm A}/\Omega_{\mathrm p}=0.001$, and isotropic temperatures for both species. The numerical damping rates for the IA mode and for the NP mode are equal at $\beta_{\parallel \mathrm p}\approx 0.3$. }
\label{fig_damping_rates}
\end{figure}

In Section~\ref{sect_slow_modes}, we discuss the nature of kinetic slow modes  at different $\beta_{\parallel\mathrm p}$ and illustrate the fluctuations in $R_{\mathrm p}$ and $\beta_{\parallel\mathrm p}$ for the relevant modes.
Section~\ref{sect_magnetosonic} explains the isotropization mechanism and quantitatively describes the effects of compressive modes on the equilibrium anisotropy. We conclude our presentation in Section~\ref{sect_conclusions}. In  Appendices \ref{app_kin}--\ref{app_lower_gamma}, we present the mathematical and numerical framework for our kinetic analysis, illustrate the fluctuating-anisotropy effect using double-adiabatic MHD, discuss the efficiency of pitch-angle scattering by FM/W waves, and analyze the dependence of the isotropization mechanism on the assumed maximum allowable growth rate of the driven microinstabilities.

\section{Slow Modes in Kinetic Theory}\label{sect_slow_modes}

For simplicity, we model the large-scale compressions as linear waves in a hot, collisionless proton--electron plasma using Vlasov--Maxwell theory. Our use of linear theory is motivated in part by the argument that strongly turbulent fluctuations in a plasma retain certain properties associated with linear modes \citep{klein12,salem12,chen13,howes14}. 
Observations in the solar wind show that $\delta n_{\mathrm p}$ and $\delta |\vec B|$ are anticorrelated \citep{belcher71,bavassano89,tu95,chernyshov08,yao11,howes12,klein12,kiyani13},
 and we consequently take the large-scale compressions to be solutions to the hot-plasma dispersion relation for which $\delta n_{\mathrm p}$ and $\delta |\vec B|$ are anticorrelated. We refer to such solutions as ``kinetic slow modes.''  We use the approximation $\delta |\vec B|\approx \delta B_{\parallel}=\delta \vec B\cdot  \hat{\vec b}$.

There are two types of kinetic slow modes: ion-acoustic (IA) waves and nonpropagating (NP) modes \citep{howes06,schekochihin09}.  We first describe these modes analytically using a set of approximations that does not apply in its entirety to slow modes in the solar wind. However, we will later relax these approximations and describe these waves numerically for parameters that are relevant to the solar wind.

For an isotropic plasma with proton and electron temperatures $T_{\mathrm p}$ and $T_{\mathrm e}$, the IA wave can be described analytically in the gyrokinetic approximation given the assumptions that $k_{\perp}\rho_{\mathrm p}\ll1$, $\omega_{\mathrm r}\ll \Omega_{\mathrm p}$, $T_{\mathrm e}\gg T_{\mathrm p}$, and $\beta_{\mathrm p}\ll 1$, where $\rho_{\mathrm p}$ is the proton gyroradius, $\omega_{\mathrm r}$ is the real part of the frequency, and $\beta_{\mathrm p}\equiv 8\pi n_{\mathrm p}k_{\mathrm B}T_{\mathrm p}/B_0^2$. In these limits, the IA dispersion relation yields \citep{howes06}
\begin{equation}\label{omegaIA}
\omega_{\mathrm r}=k_{\parallel}c_{\mathrm s}
\end{equation}
and
\begin{equation}\label{gammaIAW}
\gamma=-|k_{\parallel}|c_{\mathrm s}\sqrt{\pi}\left(\frac{T_{\mathrm e}}{2T_{\mathrm p}}\right)^{3/2}e^{-T_{\mathrm e}/2T_{\mathrm p}},
\end{equation}
where $\gamma$ is the imaginary part of the frequency and \citep[e.g.,][]{stix92,gary93,narita15}
\begin{equation}\label{csIA}
c_{\mathrm s}\equiv \sqrt{\frac{k_{\mathrm B}T_{ \mathrm e}}{m_{\mathrm p}}}.
\end{equation}
A comparison of Equations~(\ref{omegaIA}) and (\ref{csIA}) with numerical solutions to the hot-plasma dispersion relation reveals that Equation~(\ref{omegaIA}) provides a reasonably accurate approximation of the numerical solutions at $\beta_{\mathrm p}\lesssim 1$ even when $T_{\mathrm e}\approx T_{\mathrm p}$.

The NP mode can be described analytically in the gyrokinetic approximation given the assumptions that $k_{\perp}\rho_{\mathrm p}\ll 1$, $\omega_{\mathrm r}\ll \Omega_{\mathrm p}$, and $\beta_{\parallel\mathrm p}\gg 1$. As shown by \citet{kunz15}, the NP dispersion relation in these limits yields
\begin{equation}\label{gammaNPM}
\gamma\approx -\frac{|k_{\parallel}|v_{\mathrm A}}{R_{\mathrm p}^2\sqrt{\pi \beta_{\parallel\mathrm p}}}\left(1-\beta_{\perp\mathrm p}\Delta_{\mathrm p}-\beta_{\perp\mathrm e}\Delta_{\mathrm e}\right),
\end{equation}
where $\Delta_{j}\equiv R_j-1$, $\beta_{\perp j}\equiv \beta_{\parallel j} T_{\perp j}/T_{\parallel j}$, and $v_{\mathrm A}\equiv B_0/\sqrt{4\pi n_{\mathrm p}m_{\mathrm p}}$ is the proton Alfv\'en speed.\footnote{Interestingly, anisotropy can drive the NP mode unstable. The instability criterion is $\gamma>0$ in Equation~(\ref{gammaNPM}), which leads to the mirror-mode instability criterion $\beta_{\perp\mathrm p}\Delta_{\mathrm p}+\beta_{\perp\mathrm e}\Delta_{\mathrm e}>1$ \citep{kunz15}.}
We note that the analytical approximation in Equation~(\ref{gammaNPM}) does not rely on the assumption $T_{\mathrm e}\gg T_{\mathrm p}$.

In Figure~\ref{fig_damping_rates}, we compare our analytical expressions for $\gamma$ in Equations~(\ref{gammaIAW}) and (\ref{gammaNPM}) to numerical solutions of the hot-plasma dispersion relation obtained with our  New Hampshire Dispersion-relation Solver (NHDS) code \citep{verscharen13a}. We denote the angle between $\vec k$ and $\vec B_0$ as $\theta$ and set $\theta=88^{\circ}$ and $k_{\parallel}v_{\mathrm A}/\Omega_{\mathrm p}=0.001$ in Figure~\ref{fig_damping_rates}. We see that the analytical results and the numerical results agree well for the NP mode at large $\beta_{\parallel \mathrm p}$. There is a significant discrepancy between Equation~(\ref{gammaIAW}) and the numerical solutions since the assumption $T_{\parallel \mathrm e}\gg T_{\parallel\mathrm p}$ is not satisfied in the numerical solution. Using this parameter set, the NP mode has a higher damping rate than the IA mode at $\beta_{\parallel \mathrm p}\lesssim 0.3$  and a lower damping rate than the IA mode at $\beta_{\parallel \mathrm p}\gtrsim 0.3$.

If the compressive fluctuations in the solar wind were simply freely decaying slow waves, then at each $\beta_{\parallel\mathrm p}$, the dominant component of the compressions would correspond to the least damped kinetic slow mode -- i.e., the IA mode at low $\beta_{\parallel\mathrm p}$ and the NP mode at high $\beta_{\parallel\mathrm p}$. In the solar wind, however, nonlinear interactions among noncompressive fluctuations generate compressions, presumably exciting both IA and NP modes, at least to some degree, at all $\beta_{\parallel\mathrm p}$. We thus assume that a nonnegligible fraction of the compressions in the solar wind are in the form of IA modes. The observation of slow modes in the solar wind indicates that Landau damping does not suppress slow-mode turbulence completely. Highly oblique propagation is one possible explanation for the presence of slow modes. An `anti-phase-mixing' effect due to the turbulent background is an alternative explanation \citep[for details, see][]{schekochihin16}. In addition, proton beams can change the velocity-space gradient of the background distribution at the resonant velocity so that the damping rate is  smaller than in the Maxwellian case \citep{bavassano04}.

Like other fluctuating quantities, $R_{\mathrm p}$ and $\beta_{\parallel\mathrm p}$ fluctuate in a plasma wave. 
If the amplitude $\delta |\vec B|/B_0$ of the compressive IA component is large enough, even an initially isotropic plasma can become so anisotropic that it crosses a threshold for an anisotropy-driven instability.\footnote{We concentrate on ion-anisotropy-driven microinstabilities only and neglect electron-driven microinstabilities.} Because the microinstabilities grow on length scales and timescales much smaller than the length scales and timescales of the compressive fluctuations at the outer scale of the turbulence, we treat the plasma that is perturbed by the large-scale fluctuations as effectively uniform and static for the purposes of analyzing the microinstabilities \citep[see][]{marsch11,verscharen11}.

\begin{figure}
\epsscale{1.2}
\plotone{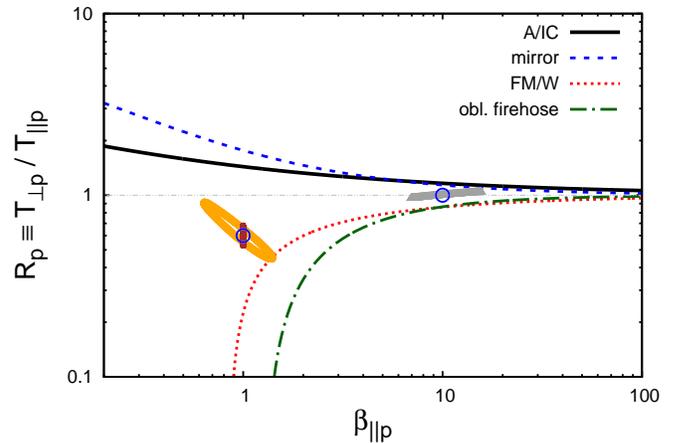}
\caption{Hodogram of a plasma parcel in kinetic theory. The orange curve shows the hodogram for the IA mode with $\delta |\vec B|/B_0=0.025$, and the brown curve shows the hodogram for the NP mode with the same $\delta |\vec B|/B_0$. This value for $\delta|\vec B|/B_0$ is the amplitude for which the IA  mode crosses the threshold of the FM/W instability. We use $\theta=88^{\circ}$, $k_{\parallel}v_{\mathrm A}/\Omega_{\mathrm p}=0.001$, $R_{0\mathrm p}=0.6$,  $T_{\parallel 0 \mathrm p}=T_{\parallel 0\mathrm e}$, and $\beta_{\parallel 0\mathrm p}=1$ for both modes.
The gray curve shows the hodogram for the NP mode with $R_{0\mathrm p}=1$, $\beta_{\parallel 0\mathrm p}=10$, and $\delta |\vec B|/B_0=0.2$. The other lines show isocontours of constant maximum growth rate $\gamma_{\mathrm m}=10^{-3}\Omega_{\mathrm p}$ for the four different anisotropy-driven instabilities under consideration. The blue circles mark the points ($\beta_{\parallel 0\mathrm p}$, $R_{0\mathrm p}$). The maximum growth rate is defined as the largest value of the instability growth rate at any $\vec k$.
 }
\label{fig_bale_plot_kinetic}
\end{figure}

The IA mode and the NP mode ``transport'' the plasma through the $\beta_{\parallel\mathrm p}$-$R_{\mathrm p}$ plane in qualitatively different ways. We illustrate this in Figure~\ref{fig_bale_plot_kinetic}, which shows  hodograms for both modes. For this figure, we calculate the fluctuating values of both $R_{\mathrm p}$ and $\beta_{\parallel\mathrm p}$ using the technique described in Appendix~\ref{app_kin} for one full wave period (for the IA mode we scan in time and for the NP mode we scan in space) and follow the plasma parcel in parameter space.
In the case of the IA mode, an increase in $R_{\mathrm p}$ coincides with a decrease in $\beta_{\parallel \mathrm p}$, leading to hodograms that extend  from the upper left to the lower right. For the NP mode, this behavior is different. For example, at $\beta_{\parallel 0\mathrm p}\approx 1$ and $R_{0\mathrm p}\approx 0.6$, $\beta_{\parallel\mathrm p}$ barely changes as $R_{\mathrm p}$ fluctuates, while for $\beta_{\parallel0\mathrm p}\approx 10$ and $R_{0\mathrm p}\approx 1$ (see the gray line in Figure~\ref{fig_bale_plot_kinetic}), $R_{\mathrm p}$ barely changes as $\beta_{\parallel\mathrm p}$ fluctuates, and the hodogram is nearly parallel to the FM/W threshold.  Another difference between the two modes is that the fluctuations in $R_{\mathrm p}$ and in $\beta_{\parallel\mathrm p}$ are significantly smaller in the NP mode than in the IA mode for a given amplitude $\delta |\vec B|/B_0$ in the parameter range that we explore.
IA modes are, for these reasons, more effective than NP modes at transporting the plasma across the FM/W threshold. Much of our subsequent analysis will thus focus on IA modes as the modes relevant for the isotropization mechanism. The fluctuating-anisotropy effect associated with NP modes may be effective at transporting the plasma across the A/IC threshold, which can be important in other contexts such as accretion flows. 

Figure~\ref{fig_bale_plot_kinetic} also shows that the IA mode exhibits a phase shift between $\beta_{\parallel\mathrm p}$ and $R_{\mathrm p}$ so that the hodogram is an oval. This effect is negligible for the NP mode. Double-adiabatic MHD provides an intuitive understanding of the fluctuating-anisotropy effect as we show in Appendix~\ref{app_MHD}. The IA mode and the slow mode in double-adiabatic MHD exhibit similar phasing (see Figure~\ref{fig_bale_plot_slow} in Appendix \ref{app_MHD}).

\section{Limits on Temperature Anisotropy from the Fluctuating-anisotropy Effect}\label{sect_magnetosonic}

As the solar wind expands out to 1 au, $\beta_{\parallel 0 {\rm p}}$ increases and $R_{0\rm p}$ decreases toward the FM/W instability threshold, denoted $R_{\rm crit}(\beta_{\parallel \rm p})$. However, before $R_{0 {\rm p}}$ reaches $R_{\rm crit}$, the fluctuating value of $R_{\rm p}$ is driven below $R_{\rm crit}$ by large-wavelength, compressive, IA fluctuations. Let us suppose that IA fluctuations would cause $R_{\rm p}$ to oscillate approximately as $R_{\rm p} = R_{0 \rm p} + (\Delta R) \cos(\omega t)$ in the rest frame of some particular plasma parcel if FM/W waves could be ignored, where $\Delta R > 0$ and $R_{0 \rm p} - \Delta R < R_{\rm crit}$. Then, accounting for the FM/W instability, when $R_{\rm p}$ first drops below $R_{\rm crit}$, FM/W waves grow and cause pitch-angle scattering of the protons. This scattering maintains the condition $R_{\rm p} \simeq R_{\rm crit}$ in a self-regulating manner, because the growth rate of the FM/W waves is a rapidly increasing function of $R_{\rm crit} - R_{\rm p}$ when $R_{\rm p} < R_{\rm crit}$. As the compression from the IA wave continues and $(\Delta R)\cos(\omega t)$ continues dropping toward its minimum value of $-\Delta R$, the ongoing scattering from FM/W waves increases $R_{0\rm p}$ so that $R_{0 \rm p } + (\Delta R )\cos(\omega t)$ remains~$\simeq R_{\rm crit}$. Eventually, when the compression reaches its peak and $\cos(\omega t) = -1$, the pitch-angle scattering causes $R_{0 \rm p}$ to reach a value $ \simeq R_{\rm crit} + \Delta R$. As the IA wave continues to oscillate, $R_{\rm p}$ oscillates about its new average value. If the oscillation period is short compared to the expansion time~$\sim r/U$, where $r$ is the heliocentric distance and $U$ is the solar-wind speed, then $R_{ {\rm p}}$ just barely reaches the FM/W instability threshold when the IA wave reaches its next maximum compression. As the plasma parcel travels farther from the Sun, solar-wind expansion continues to drive $R_{0 {\rm p}}$ toward smaller values, but the above ``fluctuating-anisotropy effect'' repeats, maintaining $R_{0 \rm p}$ at a distance~$\simeq (\Delta R)$ away from $R_{\rm crit}$. If the amplitude of the IA mode is large enough, this effect causes $R_{0 \rm p}$ to be close to~1, even when the plasma is nearly collisionless. 

The main assumption that we have made in the above discussion is that the FM/W waves that are excited when $R_{\rm p}$ drops just below $R_{\rm crit}$ represent only a small fraction of the energy of the driving IA oscillation. If this were not the case, the growth of these FM/W waves would simply damp out the IA wave, leaving $R_{\rm p} \simeq R_{\rm crit}$. Our assumption amounts to taking the FM/W pitch-angle scattering process to be efficient, in the sense that a small amount of FM/W energy is enough to maintain $R_{\rm p} \simeq R_{\rm crit}$ in the presence of the IA compression. We estimate the validity of this assumption in Appendix~C.

Figure~\ref{fig_background_ani_kinetic} shows the minimum value of $R_{0\mathrm p}$ that the plasma can reach at a given amplitude $\delta |\vec B|/B_0$ of the large-scale IA wave such that the oscillating value of $R_{\mathrm p}$ just reaches the FM/W instability threshold, which we define as the value of $R_{\mathrm p}$ at each $\beta_{\parallel \mathrm p}$ at which the maximum FM/W growth rate $\gamma_{\mathrm m}$ is $10^{-3}\Omega_{\mathrm p}$. We use the methods described in Appendix~\ref{app_kin} to create this figure.
As an example, for the case in which $\beta_{\parallel 0\mathrm p}=1$ and $\delta |\vec B|/B_0=0.04$, the fluctuating-anisotropy effect causes $R_{0\mathrm p}$ to be $\simeq 0.9$ as illustrated by the orange curve in Figure~\ref{fig_background_ani_kinetic}. Without the compressive fluctuations, the plasma would be able to reach values of $R_{0\mathrm p}$ around 0.2 before it triggers the FM/W instability.  The larger the amplitude is, the more efficiently the compressive IA fluctuations reduce the average temperature anisotropy and counteract the generation of anisotropy due to the expansion of the solar wind. 
Figure~\ref{fig_background_ani_kinetic} shows that, for IA fluctuations with a fixed $\delta |\vec B|/B_0$, this isotropization mechanism becomes more efficient as $\beta_{\parallel 0\mathrm p}$ increases, since the threshold value of $R_{\mathrm p}$ for the FM/W instability approaches 1. We show the same plots as Figure~\ref{fig_background_ani_kinetic} in Appendix~\ref{app_lower_gamma} for a different assumed value of $\gamma_{\mathrm m}$.

\begin{figure}
\includegraphics[width=\columnwidth]{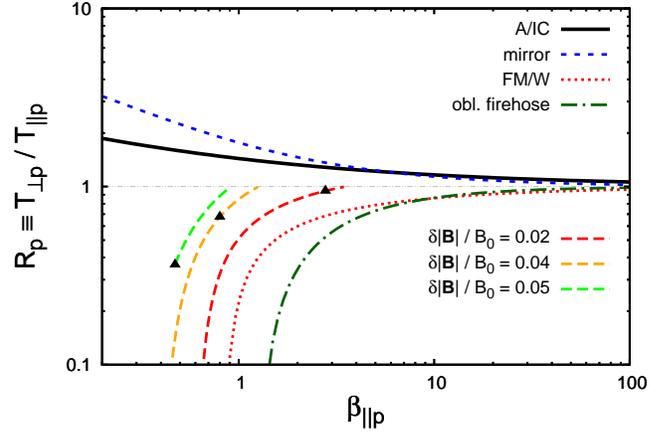}
 \caption{\label{fig_background_ani_kinetic} Permitted values of $R_{0 \mathrm p}$ after the reduction of the plasma anisotropy by the combined action of microinstabilities and compressive large-wavelength fluctuations. The colored dashed lines show the value of $R_{0 \mathrm p}$ for which the oscillating value of $R_{\mathrm p}$ barely reaches the FM/W instability threshold  ($\gamma_{\mathrm m}=10^{-3}\Omega_{\mathrm p}$) given the assumed value of $\delta |\vec B|/B_0$. In addition, we show the thresholds of the four anisotropy instabilities with $\gamma_{\mathrm  m}=10^{-3}\Omega_{\mathrm p}$. The black triangles show the point on each curve above which the IA mode with the given amplitude causes the plasma to cross  the thresholds for both the FM/W and A/IC modes.}
\end{figure}

At sufficiently large $\delta |\vec B|/B_0$, the fluctuating value of $R_{\mathrm p}$ crosses both the FM/W and A/IC thresholds. The pitch-angle scattering experienced by the protons then alternates between increasing $R_{0\mathrm p}$ and decreasing $R_{0\mathrm p}$, and $R_{\mathrm p}$ oscillates about a time-averaged value of order unity.  This double-instability regime arises along the colored dashed curves in Figure~\ref{fig_background_ani_kinetic}  above the filled black triangles. If the entire curve lies above its associated black triangle (e.g., the green line in Figure~\ref{fig_background_ani_kinetic}), every point along the curve is in the double-instability regime. We do not extend the green curve down to $R_{\mathrm p}=0.1$ because the plasma reaches $\beta_{\parallel\mathrm p}<0$ or $R_{\mathrm p}<0$ at some point during the IA oscillation for $\delta |\vec B|/B_0=0.05$ and $R_{0\mathrm p}\lesssim0.4$.

%We avoid this complication by restricting our analysis to combinations of $\delta |\vec B|/B_0$, $\beta_{\parallel 0\mathrm p}$, and $R_{0\mathrm p}$ for which only one instability threshold is crossed, which is why several of the curves in Figure~\ref{fig_background_ani_kinetic} terminate at some maximum value of $R_{0\mathrm p}$.

\section{Discussion and Conclusions}\label{sect_conclusions}

In this paper, we explore how long-wavelength compressive fluctuations and short-wavelength temperature-anisotropy instabilities work together to isotropize a weakly collisional plasma such as the solar wind.  As the solar wind flows away from the Sun in the inner heliosphere, solar-wind expansion acts to increase the average value of~$\beta_{\parallel \rm p}$ and decrease the average value of the protons' perpendicular-to-parallel temperature ratio~$R_{\rm p}$ (denoted $\beta_{\parallel 0 {\rm p}}$ and~$R_{0 {\rm p}}$, respectively). In addition, fluctuations in $B$ and $n_{\rm p}$ on timescales $\gg \Omega_{\rm p}^{-1}$ lead to fluctuations in~$\beta_{\parallel \rm p}$ and~$R_{\rm p}$ about their average values. If the fluctuating value of~$R_{\rm p}$ within some plasma parcel crosses the FM/W instability threshold~$R_{\rm crit}$, FM/W instabilities grow rapidly and cause pitch-angle scattering of the protons. This pitch-angle scattering prevents the instantaneous value of~$R_{\rm p}$ from dropping much below~$R_{\rm crit}$ and, we argue, simultaneously raises the time-averaged value of $R_{\rm p}$ within that plasma parcel. Averaged over several periods and/or wavelengths of the large-scale compressive fluctuations, this process maintains~$R_{0 {\rm p}}$ at a ``distance'' from the FM/W instability threshold in the $\beta_{\parallel {\rm p}}$-$R_{\rm p}$ plane. This distance increases with increasing $\delta |\vec B|/B_0$. In essence, we argue that $R_{0 {\rm p}}$ takes on that value for which the fluctuating value of $R_{\rm p}$ just barely crosses the FM/W threshold.

To analyze this ``fluctuating-anisotropy'' effect quantitatively, we model the large-scale compressive fluctuations as linear ion-acoustic waves, whose properties we obtain using numerical solutions to the hot-plasma dispersion relation for a collisionless plasma, supplemented by the mathematical and numerical framework developed in Appendix~\ref{app_kin}. Figure~\ref{fig_background_ani_kinetic} displays the central result of this work, showing the lower limits set on $R_{0 {\rm p}}$ by the fluctuating-anisotropy effect 
for a variety of assumed $\delta |\vec B|/B_0$ values. This figure shows that for $\delta |\vec B|/B_0\simeq 0.04$ and for $\beta_{\parallel \rm p} \gtrsim 0.7$ (typical values for the near-Earth solar wind), the fluctuating-anisotropy effect increases $R_{0 {\rm p}}$ to values between $\simeq 0.6$ and~1, well above the FM/W instability threshold and broadly consistent with solar-wind observations.\footnote{Because of the fluctuations in $R_{\rm p}$ and $\beta_{\parallel {\rm p}}$ induced by long-wavelength compressive fluctuations, plasma measurements taken with a cadence that is short compared to the periods of the compressive waves will exhibit a broad distribution of $R_{\rm p}$ and $\beta_{\parallel \rm p}$ values.}. Nonlinear effects in the solar wind may alter the phase relations between $n_{\rm p}$, $T_{\perp \rm p}$, and $T_{\parallel \rm p}$, leading to some modification of the curves in Figure~\ref{fig_background_ani_kinetic} The quantification of such modifications, however, is beyond the scope of this work. 

% In the solar wind near Earth, given plausible values of~$\delta |\vec B|/B_0$  \citep[i.e., $\delta |\vec B|/B_0\sim 0.05-0.1$,][]{bavassano04}, this distribution will extend up to the instability thresholds but be peaked near~$R_{\rm p} = 1$, consistent with observations (\citealp{kasper02}; \citealp{hellinger06}; \citealp{bale09}; \citealp{maruca12}; see also \citealp{servidio14}). 

Because $\delta |\vec B|/B_0$ need only be $\simeq 0.04$ for the IA-wave component of the compressive fluctuations, whereas $\delta |\vec B|/B_0$ is typically $\simeq 0.1$ in the solar wind, the IA waves only need to account for a minority fraction of the compressive energy in order for the fluctuating-anisotropy effect to explain the observed values of $R_{0\mathrm p}$. The observation of pressure-balanced structures (PBSs) in the solar wind \citep{burlaga70,vellante87,burlaga90,zank90, marsch93,tu94,ghosh98, bavassano04,yao11,yao13,yao13a,narita15} suggests that IA modes indeed account for only a minority of the total compressive energy, because IA waves perturb the total pressure, whereas NP modes are associated with approximate total-pressure balance in the limit $k_{\parallel}\rightarrow 0$. Future observational studies to constrain the IA fraction in the solar wind will be important for further testing of the importance of the fluctuating-anisotropy effect in the solar wind.

If the fluctuations in $R_{\rm p}$ and $\beta_{\parallel \rm p}$ are dominated by IA fluctuations, as we have assumed, then there will be comparatively few data points in the lower left corner of Figure~\ref{fig_bale_plot_kinetic}, where both $R_{\rm p}$ and $\beta_{\parallel \rm p}$ are small, because of the anticorrelation between the fluctuations in $R_{\rm p}$ and $\beta_{\parallel \rm p}$. This could explain the observed lack of measurements in this regime in the solar wind.

We note at this point that this isotropization mechanism is relevant for all collisionless turbulent plasmas as long as the frequency of the turbulent fluctuations is small compared to $\Omega_{\mathrm p}$, and $\beta_{\parallel \mathrm p}$ and $\delta |\vec B|/B_0$ are large enough.  Astrophysical plasmas that frequently fulfill these requirements include the solar wind and low-luminosity accretion disks. 
Astrophysical shear flows, for example, can create anisotropies with $R_{0\mathrm p}>1$ \citep{kunz14,riquelme15}. In the presence of compressive fluctuations, the isotropization mechanism described here can limit this equilibrium anisotropy to values that are significantly below the thresholds for the A/IC and mirror-mode instabilities. By reducing the fraction of the time that the plasma spends at or beyond the instability threshold, this effect could reduce the growth rate of plasma instabilities that inhibit thermal conduction.

Kinetic plasma simulations such as hybrid, particle-in-cell, or Vlasov-kinetic models are capable of simulating the mechanism we describe.  For example, \citet{hellinger15a} have carried out two-dimensional expanding-box hybrid simulations of solar wind turbulence that track the evolving temperature anisotropy of the plasma. Future simulations of this type in three dimensions would incorporate both the full turbulent dynamics and the efficient pitch-angle scattering by the parallel FM/W instability, leading to an important test of the fluctuating-anisotropy effect \citep[see also][]{laveder11,servidio14,servidio15}.  Numerical simulations can also test our argument that, when large-wavelength compression causes a plasma parcel to cross the FM/W threshold, the resulting amplification of FM/W fluctuations leads to pitch-angle scattering that increases $R_{0\mathrm p}$ until the fluctuating value of $R_{\mathrm p}$ stops crossing the FM/W threshold.

It is possible that there are interesting differences in the physics of this isotropization mechanism depending on the primary velocity-space instabilities that are excited.  For example, the mirror instability largely conserves the magnetic moment unless the mirrors reach large amplitudes $\delta |\vec B|/B_0 \sim 1$, while the A/IC and FM/W instabilities lead to pitch-angle scattering even at low amplitudes \citep[e.g.,][]{kunz14,riquelme15}.

Our approach can be extended to a general ``fluctuating-moment theory'' by incorporating fluctuations in other plasma bulk parameters due to large-scale compressive fluctuations and investigating their impact on kinetic instabilities. For example, fluctuations in the relative drift speed between alpha particles and protons, as well as fluctuations in their temperature anisotropy, can lead to the excitation of beam instabilities at average drift speeds below the classical thresholds of these instabilities \citep[see][]{verscharen13, verscharen13b} and thus to limits on the beam speed that are lower than predicted by linear theory. The development of this theory is beyond the scope of this work.

\acknowledgements

We appreciate helpful discussions with Phil Isenberg and Peter Gary.
This work was supported by NASA grant NNX15AI80G, NASA grant NNX16AG81G, NSF/SHINE grant AGS-1460190, NSF/SHINE grant AGS-1258998, NSF grant PHY-1500041, and NSF grant AGS-1331355. E.Q. was supported by NSF grant AST-1333612, a Simons Investigator Award from the Simons Foundation, and the David and Lucile Packard Foundation.

\appendix

\section{Appendix A\\ Kinetic-theory Treatment of the Large-scale Compressions}\label{app_kin}

In this appendix, we first discuss the thresholds of the four anisotropy-driven instabilities under consideration and then develop the mathematical and numerical framework for our kinetic treatment. For the purpose of our analysis, we describe the fluctuations in different bulk parameters with the help of linear theory. In all of our calculations, we use $v_{\mathrm A}/c=10^{-4}$.

\subsection{A.1.~Instability Thresholds}

Previous calculations have determined fits for isocontours of constant growth rates of the A/IC, mirror, FM/W, and oblique firehose instabilities in the $\beta_{\parallel \mathrm p}$-$R_{\mathrm p}$ plane. We use fits of the form
\begin{equation}\label{Rjfit}
R_{\mathrm p}=1+\frac{a}{\left(\beta_{\parallel \mathrm p}-c\right)^b}
\end{equation}
with the fit parameters $a$, $b$, and $c$  for  constant maximum growth rates $\gamma_{\mathrm m}=10^{-2}\Omega_{\mathrm p}$, $\gamma_{\mathrm m}=10^{-3}\Omega_{\mathrm p}$, and $\gamma_{\mathrm m}=10^{-4}\Omega_{\mathrm p}$.\footnote{The value for $\gamma_{\mathrm m}$ is somewhat arbitrary; however, comparisons between observations and theory show that isocontours with $\gamma_{\mathrm m}$ between $10^{-3}\Omega_{\mathrm p}$ and $10^{-2}\Omega_{\mathrm p}$ describe accurate limits for plasma parameters in the solar wind \citep[e.g.,][]{kasper02,hellinger06,bale09,maruca12,bourouaine13}. We choose  $\gamma_{\mathrm m}=10^{-3}\Omega_{\mathrm p}$. We provide fit parameters for $\gamma_{\mathrm m}=10^{-4}\Omega_{\mathrm p}$ and $\gamma_{\mathrm m}=10^{-2}\Omega_{\mathrm p}$ since we show results for these maximum growth rates in Appendix~\ref{app_lower_gamma}.} We determine these thresholds with the numerical solvers NHDS \citep{verscharen13a} and PLUME \citep{klein15}. We show the fit parameters that result from this calculation in Table~\ref{table_fit}. \citet{maruca12} give fit parameters for $\gamma_{\mathrm m}=10^{-2}\Omega_{\mathrm p}$ using fits of the same form as Equation~(\ref{Rjfit}) for a plasma containing alpha particles. The thresholds of the A/IC, the FM/W, and the oblique firehose instabilities in the case with isotropic alpha particles are slightly larger in general than in the case without alpha particles, while the mirror-mode threshold is about equal for both cases. 

\begin{deluxetable}{lccc}
\tablecaption{Fit Parameters for Isocontours of Constant $\gamma_{\mathrm m}=10^{-2}\Omega_{\mathrm p}$, $\gamma_{\mathrm m}=10^{-3}\Omega_{\mathrm p}$, and $\gamma_{\mathrm m}=10^{-4}\Omega_{\mathrm p}$ for Use in Equation~(\ref{Rjfit}).  \label{table_fit}}
\tablehead{ \colhead{Instability} & \colhead{$a$} & \colhead{$b$} & \colhead{$c$}   }
\startdata 
&$\gamma_{\mathrm m}=10^{-2}\Omega_{\mathrm p}$\\ \hline
A/IC instability & 0.649 & 0.400 & 0 \\
Mirror-mode instability & 1.040 & 0.633 & -0.012 \\ 
FM/W instability & -0.647 & 0.583 & 0.713 \\
Oblique firehose instability & -1.447 & 1.000 & -0.148 \\ \hline  \hline
&$\gamma_{\mathrm m}=10^{-3}\Omega_{\mathrm p}$ \\ \hline 
A/IC instability & 0.437 & 0.428 & -0.003 \\
Mirror-mode instability & 0.801 & 0.763 & -0.063 \\ 
FM/W instability & -0.497 & 0.566 & 0.543 \\
Oblique firehose instability & -1.390 & 1.005 & -0.111 \\ \hline \hline
&$\gamma_{\mathrm m}=10^{-4}\Omega_{\mathrm p}$ \\ \hline 
A/IC instability & 0.367 & 0.364 & 0.011 \\
Mirror-mode instability & 0.702 & 0.674 & -0.009 \\ 
FM/W instability & -0.408 & 0.529 & 0.410 \\
Oblique firehose instability & -1.454 & 1.023 & -0.178 
\enddata
\end{deluxetable}\vspace{1pt}

\subsection{A.2.~Fluctuations in the Distribution Function}

For all fluctuating quantities $A$, we introduce the notation $A(\vec r,t)=Ae^{i\vec k\cdot \vec r-i\omega t}$ with the complex amplitude $A$. The real part of $A(\vec r,t)$ is the associated observable.
Linearizing the Vlasov equation in cylindrical coordinates ($v_{\perp}$, $v_{\parallel}$, $\phi$) leads to an expression for the first-order perturbation of the distribution function:
\begin{multline}\label{df}
\delta f_{\mathrm p}(\vec r, t)=-\frac{q_{\mathrm p}}{m_{\mathrm p}}e^{i\vec k\cdot \vec r-i\omega t}\int\limits_{0}^{+\infty }d\tau e^{i\beta}\left\{\vphantom{\frac{A}{B}}E_xU\cos\left(\phi+\Omega_{\mathrm p}\tau\right)\right.\\
\left.+E_yU\sin\left(\phi+\Omega_{\mathrm p}\tau\right)+E_z\left[\diffp{f_{0\mathrm p}}{v_{\parallel}}-V\cos\left(\phi+\Omega_{\mathrm p}\tau\right)\right]\right\}
\end{multline}
\citep{stix92}, where
\begin{equation}
\beta\equiv -\frac{k_{\perp}v_{\perp}}{\Omega_{\mathrm p}}\left[\sin\left(\phi+\Omega_{\mathrm p}\tau\right)-\sin \phi\right]+\left(\omega-k_{\parallel}v_{\parallel}\right)\tau,
\end{equation}
\begin{equation}
U\equiv \diffp{f_{0\mathrm p}}{v_{\perp}}+\frac{k_{\parallel}}{\omega}\left(v_{\perp}\diffp{f_{0\mathrm p}}{v_{\parallel}}-v_{\parallel}\diffp{f_{0\mathrm p}}{v_{\perp}}\right),
\end{equation}
\begin{equation}
V\equiv \frac{k_{\perp}}{\omega}\left(v_{\perp}\diffp{f_{0\mathrm p}}{v_{\parallel}}-v_{\parallel}\diffp{f_{0\mathrm p}}{v_{\perp}}\right),
\end{equation}
and $E_x$, $E_y$, and $E_z$ are the components of the electric-field vector.
We assume that the background distribution function is bi-Maxwellian,
\begin{equation}
f_{0\mathrm p}=\frac{1}{\pi^{3/2}w_{\perp 0\mathrm p}^2w_{\parallel 0\mathrm p}}\exp\left(-\frac{v_{\perp}^2}{w_{\perp 0\mathrm p}^2}-\frac{v_{\parallel}^2}{w_{\parallel 0\mathrm p}^2}\right),
\end{equation}
where the thermal speeds are defined as $w_{\perp 0\mathrm p}\equiv\sqrt{2k_{\mathrm B}T_{\mathrm p \perp}/m_{\mathrm p}}$ and $w_{\parallel 0\mathrm p}\equiv\sqrt{2k_{\mathrm B}T_{\mathrm p \parallel}/m_{\mathrm p}}$.
Using the Bessel-function identity
\begin{equation}
e^{-i\zeta \sin\left(\phi+\Omega_{\mathrm p}\tau\right)}=\sum\limits_{n=-\infty}^{+\infty} e^{-in\left(\phi+\Omega_{\mathrm p}\tau\right)}J_n(\zeta)
\end{equation}
for the Bessel function $J_n$ of the order $n$
allows us to simplify the integral in Equation~(\ref{df}). This yields
\begin{multline}\label{dffinal}
\delta f_{\mathrm p}(\vec r,t)=-\frac{q_{\mathrm p}}{m_{\mathrm p}}e^{i\zeta\sin\phi+i\vec k\cdot \vec r-i\omega t}\sum \limits_{n=-\infty}^{+\infty}J_n(\zeta)e^{-in\phi}\\
\times \left[E_xU\frac{a_{\mathrm p}}{\Omega_{\mathrm p}^2-a_{\mathrm p}^2}\left(i\cos\phi-\frac{\Omega_{\mathrm p}}{a_{\mathrm p}}\sin\phi\right)\right.\\
+E_yU\frac{a_{\mathrm p}}{\Omega_{\mathrm p}^2-a_{\mathrm p}^2}\left(i\sin\phi+\frac{\Omega_{\mathrm p}}{a_{\mathrm p}}\cos\phi\right)-\frac{iE_z}{a_{\mathrm p}}\diffp{f_{0\mathrm p}}{v_{\parallel}}\\
-\left.E_zV\frac{a_{\mathrm p}}{\Omega_{\mathrm p}^2-a_{\mathrm p}^2}\left(i\cos\phi-\frac{\Omega_{\mathrm p}}{a_{\mathrm p}}\sin\phi\right)\right],
\end{multline}
where
\begin{equation}
a_{\mathrm p}\equiv n\Omega_{\mathrm p}-\omega+k_{\parallel}v_{\parallel}
\end{equation}
and $\zeta\equiv k_{\perp}v_{\perp}/\Omega_{\mathrm p}$.

\begin{figure}
%\epsscale{1.2}
\begin{minipage}[h]{0.49\columnwidth}
\centering
$t=0\vphantom{\frac{1}{2}}$\\ %\vspace{-0.7cm}
\includegraphics[width=0.9\textwidth]{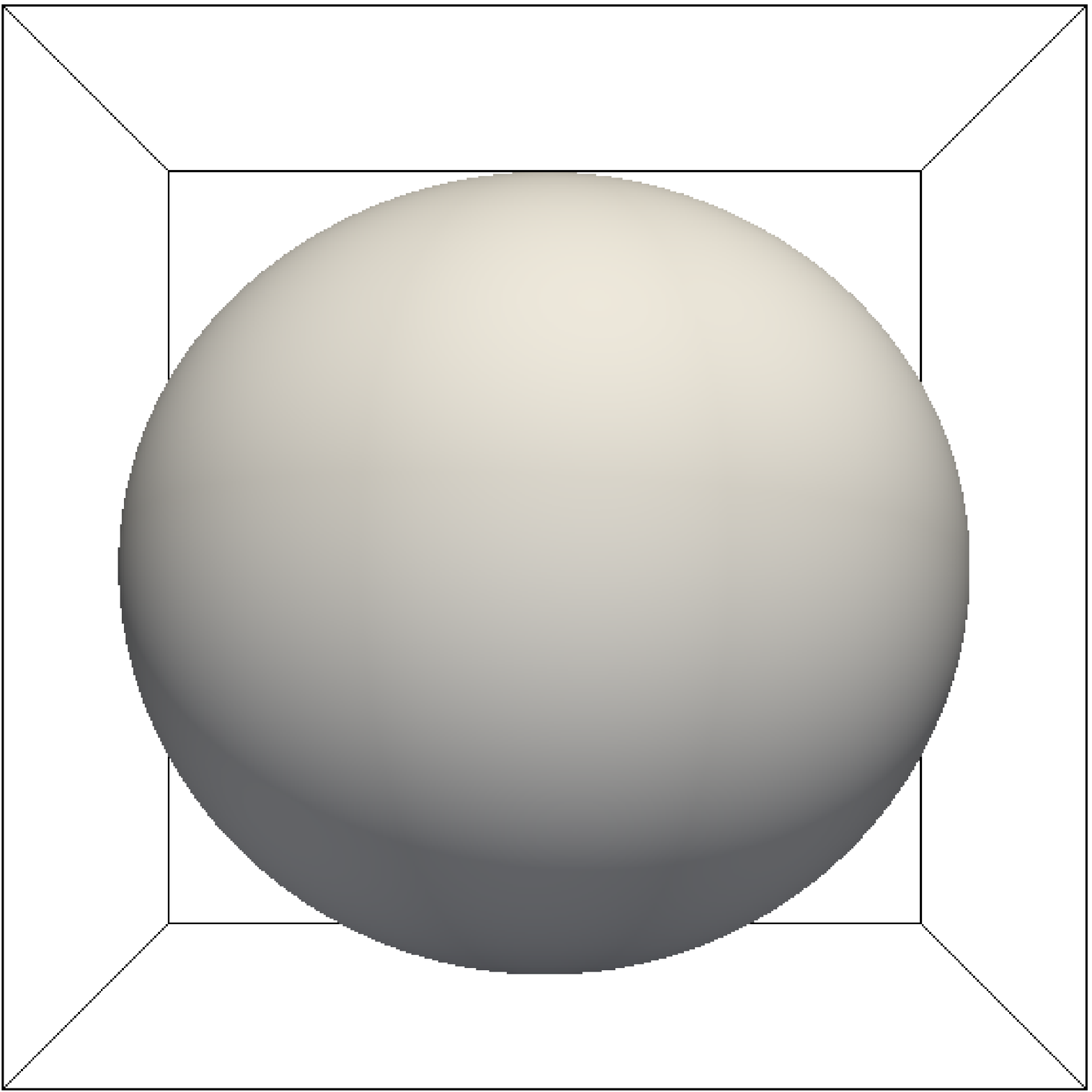}
\end{minipage}%
\begin{minipage}[h]{0.49\columnwidth}
\centering
$t=T/4\vphantom{\frac{1}{2}}$\\% \vspace{-0.7cm}
\includegraphics[width=0.9\textwidth]{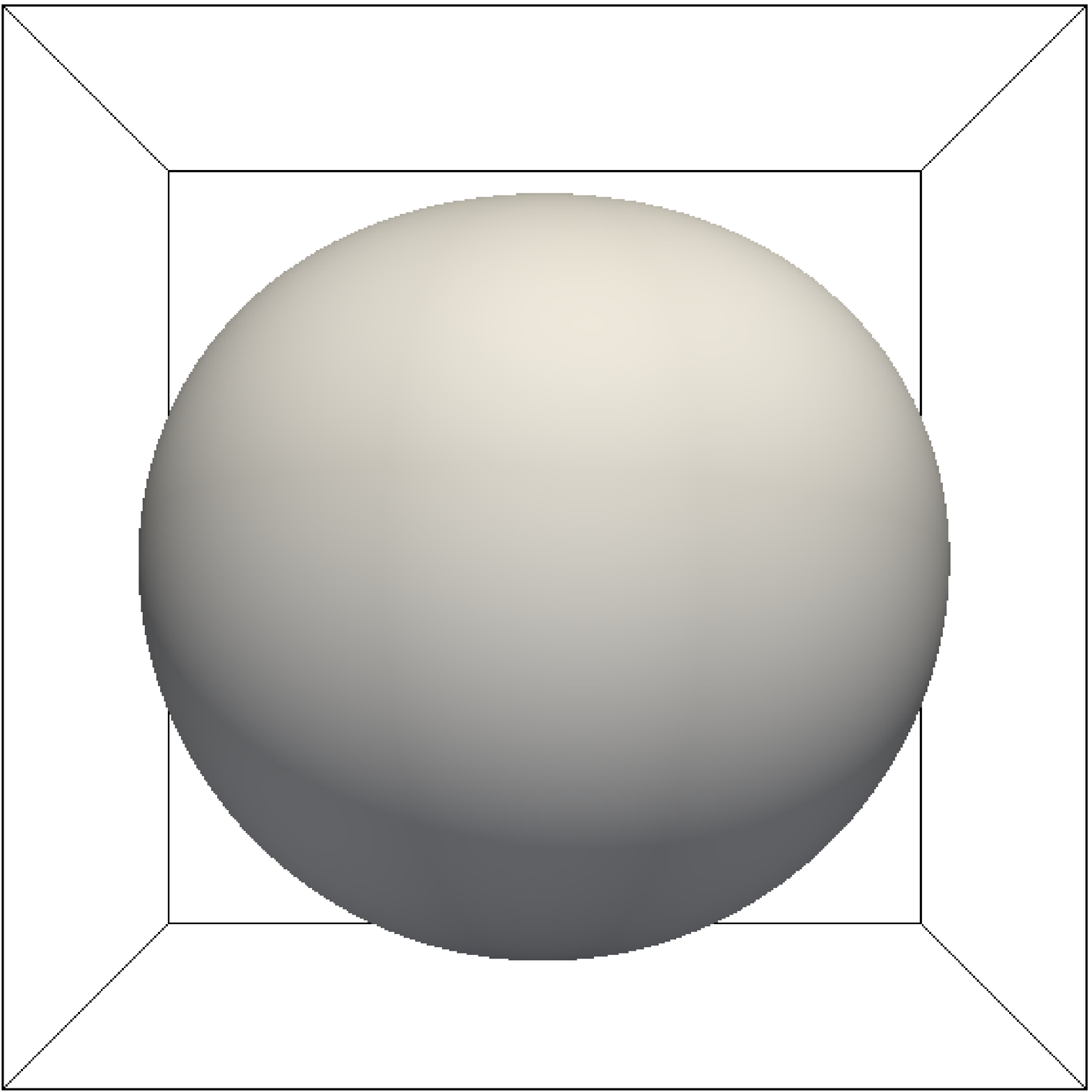}
\end{minipage}%

\begin{minipage}[h]{0.49\columnwidth}
\vspace{0.5cm}
\centering
$t=T/2\vphantom{\frac{1}{2}}$\\% \vspace{-0.7cm}
\includegraphics[width=0.9\textwidth]{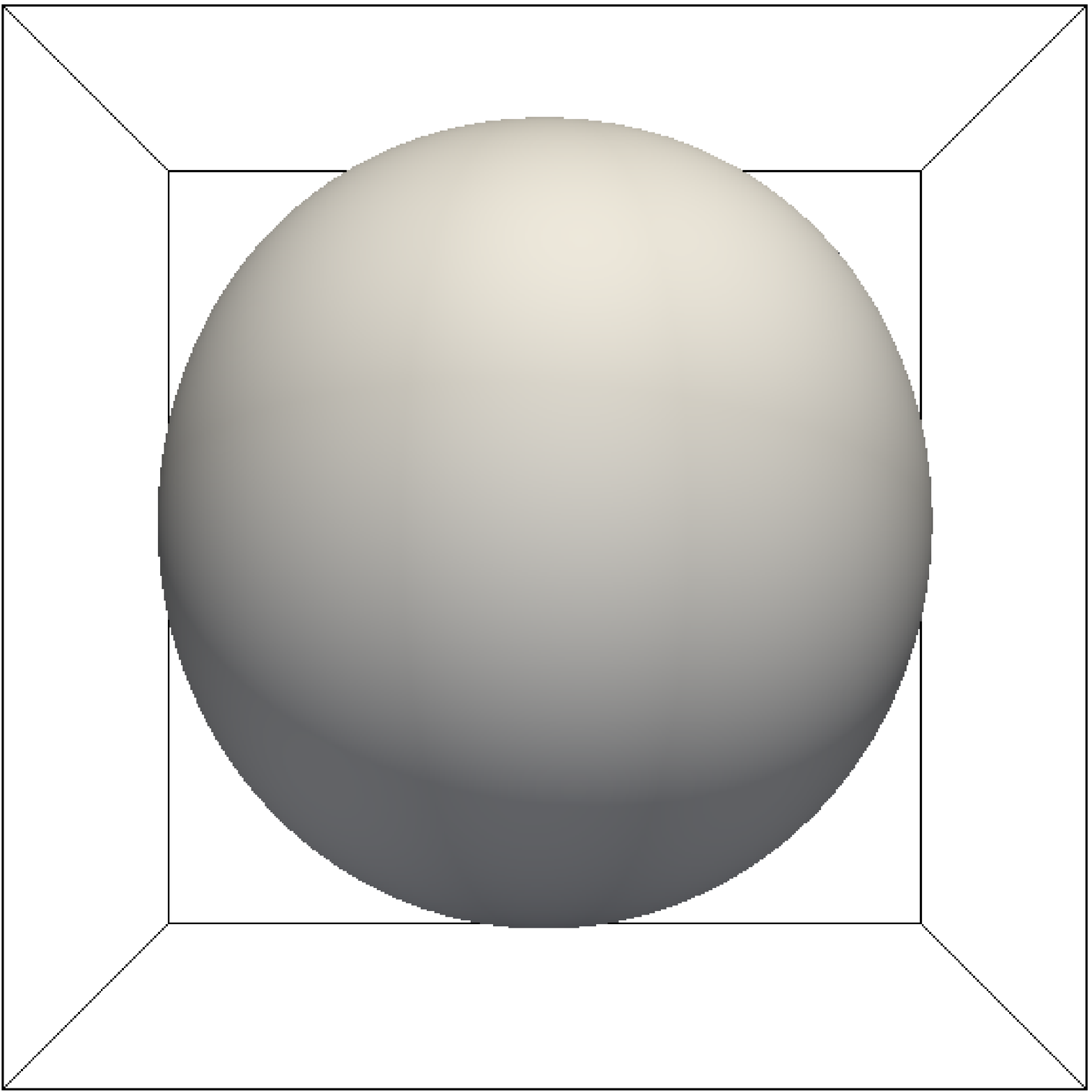}
\end{minipage}%
\begin{minipage}[h]{0.49\columnwidth}
\vspace{0.5cm}
\centering
$t=3T/4\vphantom{\frac{1}{2}}$\\% \vspace{-0.7cm}
\includegraphics[width=0.9\textwidth]{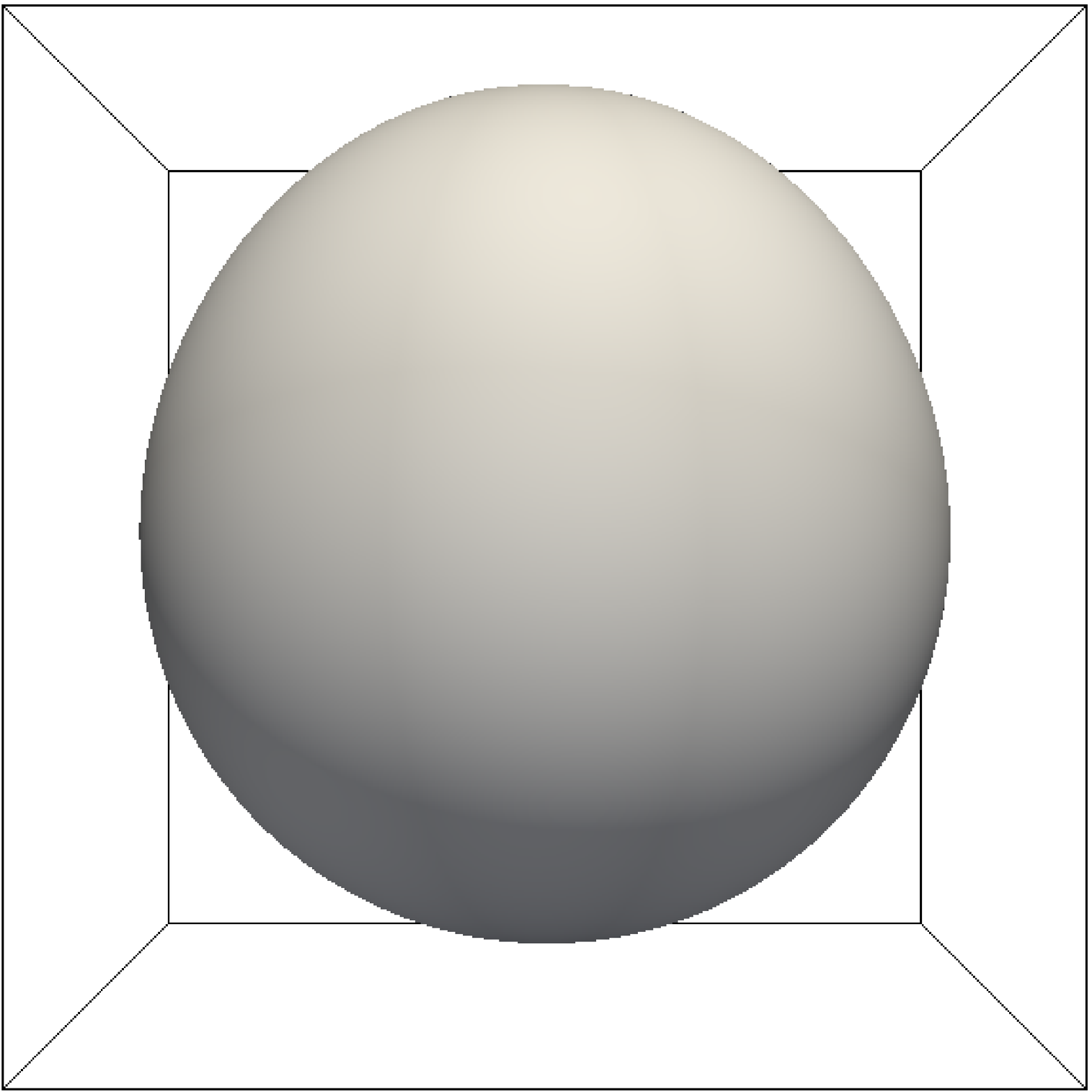}
\end{minipage}%
\vspace{0.5cm}
\caption{Isosurfaces of the proton distribution function in an IA wave with $\theta=88^{\circ}$, $\delta |\vec B|/B_0=0.1$, $\beta_{\parallel 0\mathrm p}=1$, $T_{\parallel 0 \mathrm e}=T_{\parallel 0 \mathrm p}$, and $k_{\parallel}v_{\mathrm A}/\Omega_{\mathrm p}=0.001$ at $\vec r=0$. The figure shows the distribution function at time $t=0$, $t=T/4$, $t=T/2$, and $t=3T/4$, where $T=2\pi/\omega_{\mathrm r}$ is the wave period. The direction of the background magnetic field is along the vertical axis, and the boxes span over $2v_{\mathrm A}$ in all directions. We include the effects of the large-scale compressions only and neglect the excitation of microinstabilities by the fluctuating bulk parameters and wave damping. \vspace{3pt} \\ 
(An animation of this figure is available.)}
\label{fig_slowmode}
\end{figure}

We evaluate the distribution function
\begin{equation}\label{ftotal}
f_{\mathrm p}(\vec r,t)=f_{0\mathrm p}+\delta f_{\mathrm p}(\vec r,t)
\end{equation}
using Equation~(\ref{dffinal}) with $\omega$ determined from the hot-plasma dispersion relation and with the ratios between $E_x$, $E_y$, and $E_z$ determined from the wave equation, $\vec n\times(\vec n\times \vec E)+\epsilon  \vec E=0$, where $\vec n\equiv \vec kc/\omega$ and $\epsilon$ is the hot-plasma dielectric tensor \citep{stix92}.
We show isosurfaces of the proton distribution function according to Equation~(\ref{ftotal}) for an IA mode with $\delta |\vec B|/B_0=0.1$ in Figure~\ref{fig_slowmode}. We see that both the position of the center of the distribution function along $\vec B_0$ and the temperature anisotropy (i.e., $R_{\mathrm p}$) change with the wave phase. The distribution function moves up and down along the background magnetic-field direction. It also changes its width along the field direction and across the field direction during one wave period. In the following sections, we will quantify these effects. For this and the following kinetic calculations, we choose $k_{\parallel}v_{\mathrm A}/\Omega_{\mathrm p}=0.001$, $T_{\parallel 0\mathrm e}=T_{\parallel 0 \mathrm p}$, and $\theta=88^{\circ}$ in order to avoid strong Landau damping of the IA mode. 

\subsection{A.3.~Density Fluctuations}

The numerical solver NHDS for the hot-plasma dispersion relation \citep{verscharen13a} allows us to determine the value of the complex quantity $\xi$ in
\begin{equation}\label{dnxi}
\frac{\delta n_{\mathrm p}}{n_{0\mathrm p}}=\xi \frac{\delta|\vec B|}{B_0}
\end{equation}
for any kinetic mode by using the continuity equation for electrical charge in Fourier space. The continuity equation connects the fluctuation amplitude of the density with the fluctuation amplitude of the electric field:
\begin{equation}
\frac{\delta n_{\mathrm p}}{n_0}=-i\frac{\Omega_{\mathrm p}}{\omega_{\mathrm p}^2}\vec k\cdot \chi_{\mathrm p}\frac{c\vec E}{B_0},
\end{equation}
where $\omega_{\mathrm p}\equiv\sqrt{4\pi n_{0\mathrm p}q_{\mathrm p}^2/m_{\mathrm p}}$ is the plasma frequency and $\chi_{\mathrm p}$ is the susceptibility tensor of the protons \citep{stix92}. Faraday's law and the numerically obtained relative ratios of the electric-field components then allow us to calculate $\xi$.

\begin{figure}
\epsscale{1.2}
\plotone{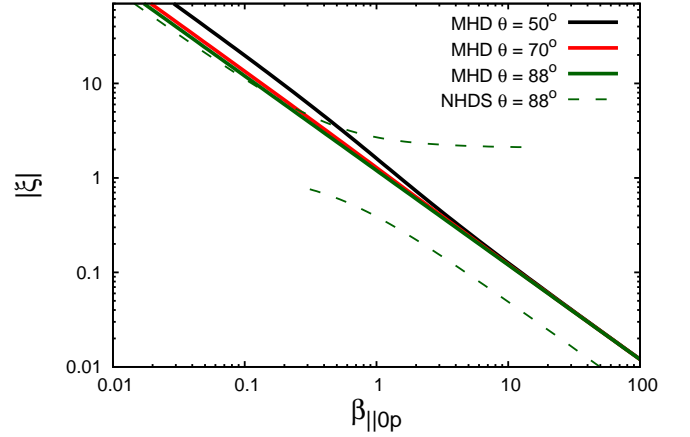}
\caption{Comparison of results for $|\xi|$ from MHD (Equation~(\ref{ximhd})) and from the solver of the kinetic hot-plasma dispersion relation NHDS for $R_{0\mathrm p}=R_{0\mathrm e}=1$ and and $T_{\parallel 0\mathrm e}=T_{\parallel 0\mathrm p}$. For the MHD solutions, we use $\kappa=5/3$. Note that $\mathrm{Re}(\xi)<0$ everywhere and $\mathrm{Im}(\xi)\ll \mathrm{Re}(\xi)$ with $\mathrm{Im}(\xi)=0$ for the MHD case. In the kinetic case, the plot shows the results for the IA mode as the curve that extends from small $\beta_{\parallel 0\mathrm p}$ to $\beta_{\parallel 0\mathrm p}=15$ and the results for the NP mode as the curve that extends from $\beta_{\parallel 0\mathrm p}=0.3$ to large $\beta_{\parallel 0\mathrm p}$. }
\label{fig_comp_MHD_kin}
\end{figure}

In Appendix~\ref{app_MHD}, we determine the value of $\xi$ in the framework of isotropic MHD. The result is given in  Equation~(\ref{ximhd}). We compare the   numerical results for $\xi$ obtained from NHDS with the results from MHD in Figure~\ref{fig_comp_MHD_kin} for the different types of slow modes. We plot $|\xi|$ as a function of $\beta_{\parallel 0\mathrm p}$ for three different angles $\theta$ in the MHD solution and for both the IA mode and the NP mode in the kinetic solution for $\theta=88^{\circ}$. We evaluate $\xi$ at the wavenumber $k_{\parallel}=0.001v_{\mathrm A}/\Omega_{\mathrm p}$.  
The absolute value of $\xi$ decreases with increasing $\beta_{\parallel 0\mathrm p}$ in all cases shown in Figure~\ref{fig_comp_MHD_kin}. This means that the density fluctuations are in general greater at smaller $\beta_{\parallel 0\mathrm p}$ when the amplitude $\delta |\vec B|/B_0$ is fixed. The different behavior for the MHD and the kinetic solutions shows the difference in polarization properties between MHD and kinetic theory. In general, the IA mode has a larger $|\xi|$ than the NP mode in this range of parameters.

\subsection{A.4.~Temperature-anisotropy Fluctuations}

The temperature anisotropy $R_{\mathrm p}$ is also a bulk-parameter polarization property that varies periodically in the presence of a slow mode. In this section, we derive the fluctuation amplitude of the temperature anisotropy in linear kinetic theory.

In our normalization ($\int f_{0\mathrm p}d^3v=1$), we find for the perpendicular (parallel) background thermal speed $w_{\perp 0 \mathrm p}$ ($w_{\parallel 0\mathrm p}$) 
\begin{equation}\label{wperp0}
w_{\perp0 \mathrm p}^2\equiv \iiint f_{0\mathrm p}v_{\perp}^3   dv_{\perp}\, dv_{\parallel}\, d\phi
\end{equation}
and
\begin{equation}
w_{\parallel 0 \mathrm p}^2\equiv2\iiint f_{0\mathrm p} v_{\parallel}^2 v_{\perp}   dv_{\perp}\, dv_{\parallel}\, d\phi.
\end{equation}
We define the fluctuations in the perpendicular (parallel) thermal speed $\delta w_{\perp \mathrm p}$ ($\delta w_{\perp \mathrm p})$ accordingly as
\begin{equation}
\delta w_{\perp \mathrm p}^2\equiv\iiint \delta f_{\mathrm p}\,v_{\perp}^3   dv_{\perp}\, dv_{\parallel}\, d\phi
\end{equation}
and
\begin{equation}\label{dwpargen}
\delta w_{\parallel \mathrm p}^2\equiv 2\iiint \delta f_{\mathrm p} v_{\parallel}^2 v_{\perp}   dv_{\perp}\, dv_{\parallel}\, d\phi.
\end{equation}
We neglect fluctuating bulk motions in the parallel and perpendicular directions since these lead to small (infinitesimal) corrections to Equations~(\ref{wperp0}) through (\ref{dwpargen}). We can then  express the fluctuating temperature ratio as
\begin{equation}\label{kinRj}
R_{\mathrm p}(\vec r,t)=\frac{w_{\perp 0\mathrm p}^2+\delta w_{\perp \mathrm p}^2(\vec r,t)}{w_{\parallel 0\mathrm p}^2+\delta w_{\parallel \mathrm p}^2(\vec r,t)}.
\end{equation}
After a long but straightforward calculation, we find
\begin{multline}\label{dwperp}
\delta w_{\perp \mathrm p}^2=-\frac{2i}{k_{\parallel}w_{\parallel 0\mathrm p}}\frac{q_\mathrm p}{m_\mathrm p}\sum\limits_{n=-\infty}^{+\infty}\left(\left\{E_x\left[I_n+\lambda_{\mathrm p}\left(I_n^{\prime}-I_n\right)\right]\frac{n\Omega_{\mathrm p}}{k_{\perp}}\right. \right.\\
\left.-iE_y\left[I_n-I_n^{\prime}-\frac{\lambda_{\mathrm p}}{2}\left(I_n+I_n^{\prime \prime}\right)+\lambda_{\mathrm p}I_n^{\prime}\right]\frac{k_{\perp}w_{\perp 0\mathrm p}^2}{\Omega_{\mathrm p}}\right\}\\
\times \left[Z_0+\left(R_{0\mathrm p}-1\right)\frac{k_{\parallel}}{\omega}Z_1\right]+E_z\left[I_n+\lambda_{\mathrm p}\left(I_n^{\prime}-I_n\right)\right]\\
\left.\times \left[\left(1-\frac{n\Omega_{\mathrm p}}{\omega}\right)R_{0\mathrm p}Z_1+\frac{n\Omega_{\mathrm p}}{\omega}Z_1\right]\vphantom{\frac{A}{B}}\right) e^{-\lambda_{\mathrm p}}
\end{multline}
and
\begin{multline}\label{dwpar}
\delta w_{\parallel \mathrm p}^2=-\frac{4i}{k_{\parallel}w_{\parallel 0\mathrm p}}\frac{q_{\mathrm p}}{m_{\mathrm p}}\sum\limits_{n=-\infty}^{+\infty}\left\{\left[E_xI_n\frac{n\Omega_{\mathrm p}}{k_{\perp}w_{\perp 0\mathrm p}^2}\right.-iE_y\left(I_n-I_n^{\prime}\right)\frac{k_{\perp}}{2\Omega_{\mathrm p}}\right]\\
\times\left[Z_2+\left(R_{0\mathrm p}-1\right)\frac{k_{\parallel}}{\omega}Z_3\right]\\
\left.+E_z\frac{I_n}{w_{\parallel 0\mathrm p}^2}\left[1+\left(\frac{1}{R_{0\mathrm p}}-1\right)\frac{n\Omega_{\mathrm p}}{\omega}\right]Z_3\right\}e^{-\lambda_{\mathrm p}},
\end{multline}
where $\lambda_{\mathrm p}\equiv k_{\perp}^2w_{\perp 0\mathrm p}^2/2\Omega_{\mathrm p}^2$, $I_n\equiv I_n(\lambda_{\mathrm p})$ denotes the modified Bessel function of order $n$, and $Z_p$ is the plasma dispersion function,\footnote{Equation~(\ref{plasmadisp}) defines $Z_p$ for $\mathrm{Im}(\omega)>0$. Analytic continuation extends this definition to $\mathrm{Im}(\omega)\le 0$ \citep[see, e.g.,][]{stix92}.}
\begin{equation}\label{plasmadisp}
Z_p\equiv-\frac{k_{\parallel}}{\sqrt{\pi}}\int \limits_{-\infty}^{+\infty}\frac{v_{\parallel}^p}{\omega-k_{\parallel}v_{\parallel}-n\Omega_{\mathrm p}}\exp\left(-\frac{v_{\parallel}^2}{w_{\parallel 0\mathrm p}^2}\right) dv_{\parallel}.
\end{equation}

In linear theory, the amplitudes of $E_x$, $E_y$, and $E_z$ can be written as a constant factor times the amplitude $\delta |\vec B|/B_0$. In principle, linear theory is only valid for infinitesimally small amplitudes, but we linearly extrapolate the fluctuations in $R_{\mathrm p}$ and $\beta_{\parallel \mathrm p }$ from the small-amplitude regime to higher amplitudes. 
We define two complex quantities $\alpha_{\perp }$ and $\alpha_{\parallel }$ as
\begin{equation}\label{dwperpsq}
\frac{\delta w_{\perp \mathrm p}^2}{v_{\mathrm A}^2}=\alpha_{\perp } \frac{\delta |\vec B|}{B_0},
\end{equation}
and
\begin{equation}
\frac{\delta w_{\parallel \mathrm p}^2}{v_{\mathrm A}^2}=\alpha_{\parallel } \frac{\delta |\vec B|}{B_0}.
\end{equation}
We find the following for the fluctuating value of $\beta_{\parallel \mathrm p}$:
\begin{equation}\label{betaalpha}
\beta_{\parallel \mathrm p}(\vec r,t)=\frac{\displaystyle 1+\xi\frac{\delta |\vec B|(\vec r,t)}{B_0}}{\displaystyle \left[1+\frac{\delta |\vec B|(\vec r,t)}{B_0}\right]^2}\left[\beta_{\parallel 0\mathrm p}+\alpha_{\parallel }\frac{\delta |\vec B|(\vec r,t)}{B_0}\right].
\end{equation}
The complex nature of the factors $\alpha_{\perp }$ and $\alpha_{\parallel }$ represents the potential phase shift between the fluctuations in $\delta |\vec B|$ and the fluctuations in $\delta w_{\perp \mathrm p}^2$ and $\delta w_{\parallel \mathrm p}^2$. This phase shift increases the difficulty in evaluating our model because $R_{\mathrm p}$ and $\beta_{\parallel \mathrm p}$ do not increase linearly in $\delta |\vec B|/B_0$. It is also the reason for the phase shift that is visible in the hodogram of the IA mode in Figure~\ref{fig_bale_plot_kinetic}.

\subsection{A.5.~Numerical Method}

We create Figure~\ref{fig_background_ani_kinetic} in the following way: We fix $k_{\parallel}v_{\mathrm A}/\Omega_{\mathrm p}=0.001$, $\theta=88^{\circ}$, and $R_{0\mathrm e}=1$. At given values for $\beta_{\parallel 0\mathrm p}=\beta_{\parallel 0\mathrm e}$ and $\delta |\vec B|/B_0$,  we solve the hot-plasma dispersion relation for $R_{0\mathrm p}=1$, which allows us to determine $\xi$, $\alpha_{\perp}$, and $\alpha_{\parallel}$ at isotropy. Then we follow the plasma parcel through $\beta_{\parallel \mathrm p}$-$R_{\mathrm p}$ space by evaluating Equations~(\ref{kinRj}) and (\ref{betaalpha}) with $\delta |\vec B|(\vec r,t)\propto \cos(T)$ over a full wave period (i.e., $T=0\dots 2\pi$) in 500 steps. If the plasma parcel stays in the FM/W-stable parameter regime (i.e., at maximum  growth rates $\gamma_{\mathrm m}$ that are $<10^{-3}\Omega_{\mathrm p}$ for the FM/W instability), we lower $R_{0\mathrm p}$ by a factor $\hat{\epsilon}=10^{1/500}$, obtain a new set of $\xi$, $\alpha_{\perp}$, and $\alpha_{\parallel}$, and follow the plasma parcel through another wave period. We repeat this procedure until the plasma reaches the isocontour for which $\gamma_{\mathrm m}=10^{-3}\Omega_{\mathrm p}$ for the FM/W instability. The present value of $R_{0\mathrm p}$ (unless $R_{0\mathrm p}\leq 0.1$ and unless $R_{\mathrm p}<0$ or $\beta_{\parallel\mathrm p}<0$ at any point during one full cycle) then gives the minimum value of the background anisotropy at the given $\beta_{\parallel 0\mathrm p}$ and $\delta |\vec B|/B_0$.

%If the plasma crosses the threshold of the FM/W instability instability but not for the A/IC instability at $R_{0\mathrm p}$, we increase $R_{0\mathrm p}$ by repeatedly multiplying it with $\hat{\epsilon}$ until the plasma just reaches the instability threshold for the FM/W instability. 

%We neglect all parameter combinations for which the plasma crosses the A/IC instability and the FM/W instability during a wave period to avoid cases in which the alternating excitation of both instabilities lead to alternating increases and reductions of $R_{0\mathrm p}$ through pitch-angle scattering from both instabilities.

\section{Appendix B\\The Fluctuating-anisotropy Effect in Double-Adiabatic MHD}\label{app_MHD}

In this section, we introduce an illustrating qualitative description of the fluctuating-anisotropy effect with the help of double-adiabatic MHD.
Here we assume that the plasma response to the large-scale modes is described by the CGL double-adiabatic equations \citep{chew56}:
\begin{equation}\label{adia1}
\diff{}{t}\left(\frac{T_{\perp \mathrm p}}{B}\right)=0
\end{equation}
and
\begin{equation}\label{adia2}
\diff{}{t}\left(\frac{B^2T_{\parallel \mathrm p}}{n_{\mathrm p}^2}\right)=0.
\end{equation}
These equations fail to account for nonzero heat flux, which is why we describe the large-scale compressions using kinetic theory in Sections~\ref{sect_slow_modes} and \ref{sect_magnetosonic}.

After some algebra, we obtain the following expressions for the differentials that describe the effect of fluctuations in $B$ and $n_{\mathrm p}$ on $R_{\mathrm p}$ and $\beta_{\parallel \mathrm p}$:
\begin{equation}\label{adiadR}
 dR_{\mathrm p}=3\frac{R_{\mathrm p}}{B} dB-2\frac{R_{\mathrm p}}{n_{\mathrm p}} dn_{\mathrm p}
\end{equation}
and 
\begin{equation}\label{adiadbeta}
 d\beta_{\parallel \mathrm p}=3\frac{\beta_{\parallel \mathrm p}}{n_{\mathrm p}} dn_{\mathrm p}-4\frac{\beta_{\parallel \mathrm p}}{B} dB.
\end{equation}
After integration, we find
\begin{equation}\label{dR}
R_{\mathrm p}=R_{0\mathrm p}\frac{\left(B/B_0\right)^3}{\left(n_{\mathrm p}/n_{0\mathrm p}\right)^2}
\end{equation}
and
\begin{equation}\label{dbeta}
\beta_{\parallel \mathrm p}=\beta_{\parallel 0\mathrm p}\frac{\left(n_{\mathrm p}/n_{0\mathrm p}\right)^3}{\left(B/B_0\right)^4},
\end{equation}
where the integration constants $R_{0\mathrm p}$ and $\beta_{\parallel 0 \mathrm p}$ define the values of $R_{\mathrm p}$ and $\beta_{\parallel \mathrm p}$ for $\delta |\vec B|=\delta n_{\mathrm p}=0$.\footnote{In addition to the following analysis of fluctuation-induced variations in $R_{\mathrm p}$ and $\beta_{\parallel \mathrm p}$, these expressions also allow us to illustrate the radial evolution of $R_{\mathrm p}$ and $\beta_{\parallel\mathrm p}$ as a consequence of the solar-wind expansion. For this demonstration, we neglect fluctuations in $B$ and $n_{\mathrm p}$ and treat $B_0$ and $n_0$ as the values of $B$ and $n$ at the distance $r$ at which $R_{\mathrm p}=R_{0\mathrm p}$ and $\beta_{\parallel\mathrm p}=\beta_{\parallel 0\mathrm p}$. The radial decrease in $B$ and $n_{\mathrm p}$ in the solar wind suggests that $R_{\mathrm p}$ decreases with radial distance $r$ from the Sun, while $\beta_{\parallel \mathrm p}$ increases with $r$ under the first-order assumption of a radial split-monopole magnetic field and a radial flow with $r$-independent bulk speed, i.e., $B\propto r^{-2}$ and $n_{ \mathrm p}\propto r^{-2}$ \citep{matteini07,matteini12,hellinger15,hellinger15a}. The split-monopole approximation breaks down at larger $r$ due to the increasing average angle between the magnetic-field direction and the radial direction. }

The parameter $\xi$ in Equation~(\ref{dnxi}) is a real scalar in MHD theory.
With its value and Equations~(\ref{dR}) and (\ref{dbeta}), we express $R_{\mathrm p}$ and $ \beta_{\parallel \mathrm p}$ as functions of $\delta |\vec B|/B_0$ alone.
For the sake of simplicity, we take our equilibrium state to be isotropic and approximate the wave-polarization properties using isotropic MHD, which yields
\begin{equation}\label{ximhd}
\xi=\frac{C_{\pm}^2}{C_{\pm}^2-\frac{\kappa}{2}\beta_{\parallel 0\mathrm p}\cos^2\theta},
\end{equation}
where
\begin{equation}\label{Cpm}
C_{\pm}^2\equiv \frac{1}{2}\left(1+\frac{\kappa}{2}\beta_{\parallel 0\mathrm p}\right)\pm \frac{1}{2}\left[\left(1+\frac{\kappa}{2}\beta_{\parallel 0\mathrm p}\right)^2-4\frac{\kappa}{2}\beta_{\parallel 0\mathrm p}\cos^2\theta\right]^{1/2}
\end{equation}
defines the phase speed of the fast (upper sign) and slow (lower sign) magnetosonic mode in units of the Alfv\'en speed $v_{\mathrm A}$, and 
where $\kappa$ is the specific heat ratio \citep[see also][]{marsch86,tu95}.

\begin{figure}
\epsscale{1.2}
\plotone{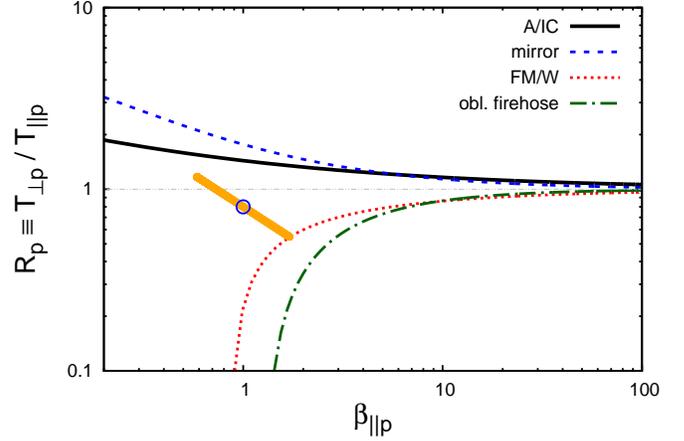}
\caption{Hodogram of a plasma parcel in slow-mode turbulence. The orange line shows the path of a plasma parcel in the $\beta_{\parallel \mathrm p}$-$R_{\mathrm p}$ plane during a full wave period for $\delta |\vec B|/B_0=0.07$, $\theta=88^{\circ}$, $R_{0\mathrm p}=0.8$, $\kappa=5/3$, and $\beta_{\parallel 0\mathrm p}=1$. The other lines show isocontours of constant maximum growth rate $\gamma_{\mathrm m}=10^{-3}\Omega_{\mathrm p}$ for four different anisotropy-driven instabilities. The blue circle marks the point ($\beta_{\parallel 0\mathrm p}$, $R_{0\mathrm p}$). In this plot, the large-scale compressions are treated using double-adiabatic MHD.}
\label{fig_bale_plot_slow}
\end{figure}

Figure~\ref{fig_bale_plot_slow} shows the hodogram of a plasma parcel in the $\beta_{\parallel \mathrm p}$-$R_{\mathrm p}$ plane for slow-mode turbulence with $\delta |\vec B|/B_0=0.07$, $R_{0\mathrm p}=0.8$, and $\beta_{\parallel 0\mathrm p}=1$, which we have constructed with Equations~(\ref{dR}) and (\ref{dbeta}) using the thresholds from Equation (\ref{Rjfit}). In addition, we show the isocontours of constant maximum growth rate $\gamma_{\mathrm m}=10^{-3}\Omega_{\mathrm p}$ for the A/IC, mirror-mode, FM/W, and oblique firehose instabilities.

\section{Appendix C\\ Estimate for the Minimum FM/W Amplitude for Effective Pitch-angle Scattering}\label{PAscatt}

For simplicity, we take all the FM/W wavevectors to be quasi-parallel ($k_x^2 + k_y^2 \ll k_{\parallel}^2$).
The general expression for the resonant pitch-angle scattering by plasma waves with right-handed polarization (i.e., FM/W modes) is then given by \citep{kennel66,stix92,marsch06}
\begin{align}\label{PAscatt}
\diffp{f_{\mathrm p}}{t}=\frac{1}{V} \int \frac{ d^3 k}{(2\pi)^3}\frac{|\tilde{\vec B}_{\mathrm F}(\vec k)|^2}{B_0^2}\frac{1}{v_{\perp}}\diffp{}{\alpha}\left( v_{\perp} \mathcal N_{\mathrm p} \diffp{f_{\mathrm p}}{\alpha}\right),
\end{align}
where
\begin{equation}
\diffp{}{\alpha}=v_{\perp}\diffp{}{v_{\parallel}}-\left(v_{\parallel}-\frac{\omega_{\mathrm{F}}}{k_{\parallel}}\right)\diffp{}{v_{\perp}},
\end{equation}
\begin{equation}
\mathcal N_{\mathrm p}\equiv \pi \frac{\Omega_{\mathrm p}^2}{\left|k_{\parallel}\right|} \delta\left(\frac{\omega_{\mathrm {F}}+\Omega_{\mathrm p}}{k_{\parallel}}-v_{\parallel} \right),
\end{equation}
$\tilde{\vec B}_{\mathrm F}(\vec k)$ is the Fourier transform of the magnetic-field fluctuations of the FM/W modes, $\omega_{\mathrm {F}}$ is the real part of the FM/W wave frequency, and $V$ is an arbitrarily large integration volume.

We assume that when the FM/W instability is excited, linear growth and nonlinear interactions between FM/W waves generate a distribution of FM/W waves over a range of $k_{\parallel}$ and $\omega_{\mathrm F}$ of order $\Omega_{\mathrm p}/v_{\mathrm A}$ and $\Omega_{\mathrm p}$, respectively, with a mean-square magnetic fluctuation given by
\begin{equation}
\delta B_{\mathrm F}^2 = \frac{1}{V} \int \frac{ d^3 k}{\left(2\pi\right)^3} \left| \tilde{\vec B}_{\mathrm F}(\vec k)\right|^2.
\label{eq:msB}
\end{equation}  
We define the one-dimensional power spectrum of the FM/W waves through the equation
\begin{equation}
P_{\mathrm F}(k_{\parallel}) \equiv \frac{1}{V} \int \frac{ dk_x\,  dk_y}{\left(2\pi\right)^3}  \left| \tilde{\vec B}_{\mathrm F}(\vec k)\right|^2,
\label{eq:defP}
\end{equation}  
in terms of which Equation~(\ref{eq:msB})  becomes
$\delta B_{\mathrm F}^2 = \int  dk_{\parallel} P_{\mathrm F}(k_\parallel)$.
Since the FM/W power is, by assumption, spread out over an interval of $k_{\parallel}$ values of width $\sim \Omega_{\mathrm p} /v_{\mathrm A}$, we obtain the estimate
\begin{equation}
P_{\mathrm F}(k_{\parallel}) \sim \frac{\delta B_{\mathrm F}^2 v_{\mathrm A}}{\Omega_{\mathrm p}}.
\label{eq:PF2} 
\end{equation}  
The interval of $k_{\parallel}$ values in which FM/W waves are excited corresponds to an interval of resonant parallel proton velocities. These velocities are moderately super-Alfv\'enic \citep[see, e.g.,][]{quest96,verscharen13b}. For parallel velocities within this velocity interval, we can use Equation~(\ref{eq:PF2}) to approximate Equation~(\ref{PAscatt}) as
\begin{equation}
\frac{\partial f_{\mathrm p}}{\partial t} \sim \frac{1}{v_{\perp}} \frac{\partial}{\partial \alpha^\prime}
\left( \nu v_{\perp} \frac{\partial f_{\mathrm p}}{\partial \alpha^\prime}\right),
\label{eq:QLD2}
\end{equation}  
where
\begin{equation}
\nu  \sim \Omega_{\mathrm p}\frac{\delta B_{\mathrm F}^2}{B_0^2}
\label{eq:defnu} 
\end{equation} 
is the effective wave pitch-angle scattering rate,
\begin{equation}
\frac{\partial }{\partial \alpha^{\prime}} = v_{\perp} \frac{\partial}{\partial v_{\parallel}}
- \left[v_{\parallel} - v_{\mathrm{ph}}\left(v_{\parallel}\right)\right] \frac{\partial }{\partial v_{\perp}},
\label{eq:defalphapr} 
\end{equation} 
and $v_{\mathrm {ph}}\left(v_{\parallel}\right)$ is the parallel phase velocity ($\omega_{\mathrm F}/k_\parallel$) of the FM/W waves that are resonant with protons whose parallel velocity is~$v_{\parallel}$.

On the other hand, the typical timescale on which the large-scale IA mode increases the temperature anisotropy is of order $1/\omega_{\mathrm r}$, where $\omega_{\mathrm r}$ is the real-part of the IA-mode frequency. Therefore, the pitch-angle diffusion is sufficient to hold the plasma at the instability threshold provided that
\begin{equation}\label{FMWamplitude}
\frac{\delta  B_{\mathrm F}^2}{B_0^2}\sim \frac{\omega_{\mathrm r}}{\Omega_{\mathrm p}}.
\end{equation}
Observations in the solar wind at 1 au show that the typical transit time for outer-scale fluctuations in the spacecraft frame is of order $\tau\sim 3\times 10^3\,\mathrm s$ \citep{alexandrova13,bruno13}. The typical proton cyclotron frequency is $\Omega_{\mathrm p}\sim 3\times 10^{-1}\,\mathrm s^{-1}$. According to Taylor's hypothesis \citep{taylor38,fredricks76}, the spatial scale of the fluctuations $\lambda$ satisfies $\lambda\sim U\tau$, where $U$ is the solar-wind speed. We set $U/c_{\mathrm s}\sim 10$ and $k_{\parallel}\sim 1/\lambda$. Therefore, we can estimate that
\begin{equation}\label{scalings}
\frac{\omega_{\mathrm r}}{\Omega_{\mathrm p}}\sim \frac{k_{\parallel}c_{\mathrm s}}{\Omega_{\mathrm p}}\sim \frac{c_{\mathrm s}}{\lambda \Omega_{\mathrm p}}\sim \frac{c_{\mathrm s}}{U}\frac{1}{\tau \Omega_{\mathrm p}} \sim 10^{-4}.
\end{equation}
for the IA waves at the outer scales of the compressive turbulence that we consider. Upon inserting Equation~(\ref{scalings}) into Equation~(\ref{FMWamplitude}), we find that the energy of the FM/W waves is much smaller than the energy of the IA waves, as we assumed in Section~\ref{sect_magnetosonic}.

\section{Appendix D\\ Limits on $R_{0\MakeLowercase{\mathrm p}}$ for $\gamma_{\MakeLowercase{\mathrm m}}=10^{-2}\Omega_{\MakeLowercase{\mathrm p}}$ and $\gamma_{\MakeLowercase{\mathrm m}}=10^{-4}\Omega_{\MakeLowercase{\mathrm p}}$}\label{app_lower_gamma}

\begin{figure}
\includegraphics[width=\columnwidth]{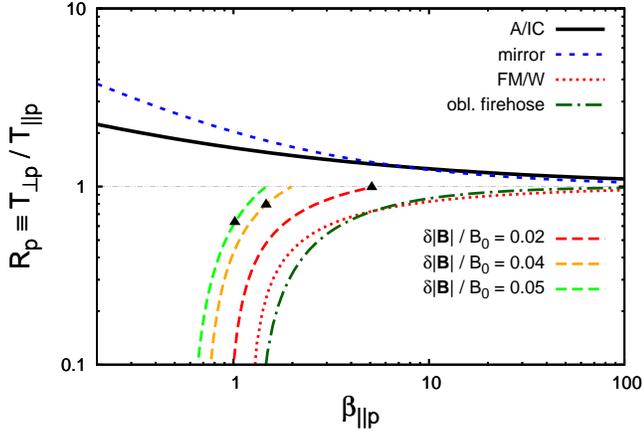}
 \caption{\label{fig_background_ani_kinetic_higher_gamma} 
Same as Figure~\ref{fig_background_ani_kinetic}, except that we use $\gamma_{\mathrm m}=10^{-2}\Omega_{\mathrm p}$.}
\end{figure}

\begin{figure}
\includegraphics[width=\columnwidth]{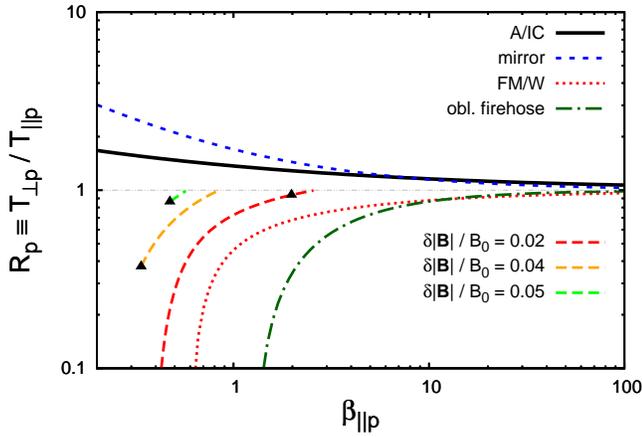}
 \caption{\label{fig_background_ani_kinetic_lower_gamma} 
Same figure as Figure~\ref{fig_background_ani_kinetic}, except that we use $\gamma_{\mathrm m}=10^{-4}\Omega_{\mathrm p}$.}
\end{figure}

We demonstrate the $\gamma_{\mathrm m}$ dependence of the isotropization mechanism in Figure~\ref{fig_background_ani_kinetic_higher_gamma}, which is the same as Figure~\ref{fig_background_ani_kinetic} except that we use $\gamma_{\mathrm m}=10^{-2}\Omega_{\mathrm p}$.
In Figure~\ref{fig_background_ani_kinetic_lower_gamma}, we show the same figure except that we use $\gamma_{\mathrm m}=10^{-4}\Omega_{\mathrm p}$.
For $\gamma_{\mathrm m}=10^{-4}\Omega_{\mathrm p}$, the values for $R_{0\mathrm p}$ are greater than the values in the case with $\gamma_{\mathrm m}=10^{-3}\Omega_{\mathrm p}$ at the same amplitude $\delta |\vec B|/B_0$. It is the opposite for $\gamma_{\mathrm m}=10^{-2}\Omega_{\mathrm p}$.  For example, plasma with $\beta_{\parallel 0\mathrm p}= 1$ and $\delta |\vec B|/B_0=0.02$ is restricted to $R_{0\mathrm p}\gtrsim 0.7$ if $\gamma_{\mathrm m}=10^{-4}\Omega_{\mathrm p}$, whereas it would be limited to $R_{0\mathrm p}\gtrsim 0.5$ when $\gamma_{\mathrm m}=10^{-3}\Omega_{\mathrm p}$ (see Figure~\ref{fig_background_ani_kinetic}). In the case with $\gamma_{\mathrm m}=10^{-2}\Omega_{\mathrm p}$ and $\beta_{\parallel 0\mathrm p}=1$,  it could reach values of $R_{0\mathrm p}$ as low as $\sim 0.1$ before the fluctuating-anisotropy effect begins to isotropize the plasma.

\bibliographystyle{aasjournal}
\bibliography{isotropization}

\begin{thebibliography}{}
\expandafter\ifx\csname natexlab\endcsname\relax\def\natexlab#1{#1}\fi

\bibitem[{{Alexandrova} {et~al.}(2013){Alexandrova}, {Chen}, {Sorriso-Valvo},
  {Horbury}, \& {Bale}}]{alexandrova13}
{Alexandrova}, O., {Chen}, C.~H.~K., {Sorriso-Valvo}, L., {Horbury}, T.~S., \&
  {Bale}, S.~D. 2013, SSRv, 178, 101

\bibitem[{{Bale} {et~al.}(2009){Bale}, {Kasper}, {Howes}, {Quataert}, {Salem},
  \& {Sundkvist}}]{bale09}
{Bale}, S.~D., {Kasper}, J.~C., {Howes}, G.~G., {et~al.} 2009, PhRvL, 103,
  211101

\bibitem[{{Bavassano} \& {Bruno}(1989)}]{bavassano89}
{Bavassano}, B., \& {Bruno}, R. 1989, JGR, 94, 11977

\bibitem[{{Bavassano} {et~al.}(2004){Bavassano}, {Pietropaolo}, \&
  {Bruno}}]{bavassano04}
{Bavassano}, B., {Pietropaolo}, E., \& {Bruno}, R. 2004, AnGeo, 22, 689

\bibitem[{{Belcher} \& {Davis}(1971)}]{belcher71}
{Belcher}, J.~W., \& {Davis}, Jr., L. 1971, JGR, 76, 3534

\bibitem[{{Bourouaine} {et~al.}(2013){Bourouaine}, {Verscharen}, {Chandran},
  {Maruca}, \& {Kasper}}]{bourouaine13}
{Bourouaine}, S., {Verscharen}, D., {Chandran}, B.~D.~G., {Maruca}, B.~A., \&
  {Kasper}, J.~C. 2013, \apjl, 777, L3

\bibitem[{{Bruno} \& {Carbone}(2013)}]{bruno13}
{Bruno}, R., \& {Carbone}, V. 2013, LRSP, 10, 2

\bibitem[{{Burlaga} \& {Ogilvie}(1970)}]{burlaga70}
{Burlaga}, L.~F., \& {Ogilvie}, K.~W. 1970, SoPh, 15, 61

\bibitem[{{Burlaga} {et~al.}(1990){Burlaga}, {Scudder}, {Klein}, \&
  {Isenberg}}]{burlaga90}
{Burlaga}, L.~F., {Scudder}, J.~D., {Klein}, L.~W., \& {Isenberg}, P.~A. 1990,
  JGR, 95, 2229

\bibitem[{{Carbone}(2012)}]{carbone12}
{Carbone}, V. 2012, SSRv, 172, 343

\bibitem[{{Chandran} {et~al.}(2011){Chandran}, {Dennis}, {Quataert}, \&
  {Bale}}]{chandran11}
{Chandran}, B.~D.~G., {Dennis}, T.~J., {Quataert}, E., \& {Bale}, S.~D. 2011,
  \apj, 743, 197

\bibitem[{{Chen} {et~al.}(2013){Chen}, {Boldyrev}, {Xia}, \& {Perez}}]{chen13}
{Chen}, C.~H.~K., {Boldyrev}, S., {Xia}, Q., \& {Perez}, J.~C. 2013, PhRvL,
  110, 225002

\bibitem[{{Chen} {et~al.}(2011){Chen}, {Mallet}, {Yousef}, {Schekochihin}, \&
  {Horbury}}]{chen11}
{Chen}, C.~H.~K., {Mallet}, A., {Yousef}, T.~A., {Schekochihin}, A.~A., \&
  {Horbury}, T.~S. 2011, \mnras, 415, 3219

\bibitem[{{Chernyshov} {et~al.}(2008){Chernyshov}, {Karelsky}, \&
  {Petrosyan}}]{chernyshov08}
{Chernyshov}, A.~A., {Karelsky}, K.~V., \& {Petrosyan}, A.~S. 2008, \apj, 686,
  1137

\bibitem[{{Chew} {et~al.}(1956){Chew}, {Goldberger}, \& {Low}}]{chew56}
{Chew}, G.~F., {Goldberger}, M.~L., \& {Low}, F.~E. 1956, Royal Society of
  London Proceedings Series A, 236, 112

\bibitem[{{Cho} \& {Vishniac}(2000)}]{cho00}
{Cho}, J., \& {Vishniac}, E.~T. 2000, \apj, 539, 273

\bibitem[{{Fredricks} \& {Coroniti}(1976)}]{fredricks76}
{Fredricks}, R.~W., \& {Coroniti}, F.~V. 1976, JGR, 81, 5591

\bibitem[{{Gary}(1993)}]{gary93}
{Gary}, S.~P. 1993, {Theory of Space Plasma Microinstabilities, Cambridge
  University Press, Cambridge, UK}

\bibitem[{{Gary} {et~al.}(2016){Gary}, {Jian}, {Broiles}, {Stevens}, {Podesta},
  \& {Kasper}}]{gary16}
{Gary}, S.~P., {Jian}, L.~K., {Broiles}, T.~W., {et~al.} 2016, JGR, 121, 30

\bibitem[{{Gary} \& {Lee}(1994)}]{gary94}
{Gary}, S.~P., \& {Lee}, M.~A. 1994, JGR, 99, 11297

\bibitem[{{Gary} {et~al.}(1998){Gary}, {Li}, {O'Rourke}, \& {Winske}}]{gary98}
{Gary}, S.~P., {Li}, H., {O'Rourke}, S., \& {Winske}, D. 1998, JGR, 103, 14567

\bibitem[{{Ghosh} {et~al.}(1998){Ghosh}, {Matthaeus}, {Roberts}, \&
  {Goldstein}}]{ghosh98}
{Ghosh}, S., {Matthaeus}, W.~H., {Roberts}, D.~A., \& {Goldstein}, M.~L. 1998,
  JGR, 103, 23705

\bibitem[{{Goldreich} \& {Sridhar}(1995)}]{goldreich05}
{Goldreich}, P., \& {Sridhar}, S. 1995, \apj, 438, 763

\bibitem[{{He} {et~al.}(2012){He}, {Tu}, {Marsch}, \& {Yao}}]{he12}
{He}, J., {Tu}, C., {Marsch}, E., \& {Yao}, S. 2012, \apjl, 745, L8

\bibitem[{{Hellinger} \& {Matsumoto}(2000)}]{hellinger00}
{Hellinger}, P., \& {Matsumoto}, H. 2000, JGR, 105, 10519

\bibitem[{{Hellinger} {et~al.}(2015){Hellinger}, {Matteini}, {Landi},
  {Verdini}, {Franci}, \& {Tr{\'a}vn{\'{\i}}{\v c}ek}}]{hellinger15a}
{Hellinger}, P., {Matteini}, L., {Landi}, S., {et~al.} 2015, \apjl, 811, L32

\bibitem[{{Hellinger} {et~al.}(2006){Hellinger}, {Tr{\'a}vn{\'{\i}}{\v c}ek},
  {Kasper}, \& {Lazarus}}]{hellinger06}
{Hellinger}, P., {Tr{\'a}vn{\'{\i}}{\v c}ek}, P., {Kasper}, J.~C., \&
  {Lazarus}, A.~J. 2006, GeoRL, 33, 9101

\bibitem[{{Hellinger} \& {Tr{\'a}vn{\'{\i}}{\v c}ek}(2008)}]{hellinger08}
{Hellinger}, P., \& {Tr{\'a}vn{\'{\i}}{\v c}ek}, P.~M. 2008, JGR, 113, 10109

\bibitem[{{Hellinger} \& {Tr{\'a}vn{\'{\i}}{\v c}ek}(2015)}]{hellinger15}
---. 2015, JPlPh, 81, 305810103

\bibitem[{{Howes} {et~al.}(2012){Howes}, {Bale}, {Klein}, {Chen}, {Salem}, \&
  {TenBarge}}]{howes12}
{Howes}, G.~G., {Bale}, S.~D., {Klein}, K.~G., {et~al.} 2012, \apjl, 753, L19

\bibitem[{{Howes} {et~al.}(2006){Howes}, {Cowley}, {Dorland}, {Hammett},
  {Quataert}, \& {Schekochihin}}]{howes06}
{Howes}, G.~G., {Cowley}, S.~C., {Dorland}, W., {et~al.} 2006, \apj, 651, 590

\bibitem[{{Howes} {et~al.}(2014){Howes}, {Klein}, \& {TenBarge}}]{howes14}
{Howes}, G.~G., {Klein}, K.~G., \& {TenBarge}, J.~M. 2014, ArXiv e-prints,
  arXiv:1404.2913

\bibitem[{{Kasper}(2002)}]{kasper02}
{Kasper}, J.~C. 2002, PhD thesis, Massachusetts Institute of Technology

\bibitem[{{Kennel} \& {Engelmann}(1966)}]{kennel66}
{Kennel}, C.~F., \& {Engelmann}, F. 1966, PhFl, 9, 2377

\bibitem[{{Kiyani} {et~al.}(2013){Kiyani}, {Chapman}, {Sahraoui}, {Hnat},
  {Fauvarque}, \& {Khotyaintsev}}]{kiyani13}
{Kiyani}, K.~H., {Chapman}, S.~C., {Sahraoui}, F., {et~al.} 2013, \apj, 763, 10

\bibitem[{{Klein} \& {Howes}(2015)}]{klein15}
{Klein}, K.~G., \& {Howes}, G.~G. 2015, PhPl, 22, 032903

\bibitem[{{Klein} {et~al.}(2012){Klein}, {Howes}, {TenBarge}, {Bale}, {Chen},
  \& {Salem}}]{klein12}
{Klein}, K.~G., {Howes}, G.~G., {TenBarge}, J.~M., {et~al.} 2012, \apj, 755,
  159

\bibitem[{{Kunz} {et~al.}(2015){Kunz}, {Schekochihin}, {Chen}, {Abel}, \&
  {Cowley}}]{kunz15}
{Kunz}, M.~W., {Schekochihin}, A.~A., {Chen}, C.~H.~K., {Abel}, I.~G., \&
  {Cowley}, S.~C. 2015, JPlPh, 81, 325810501

\bibitem[{{Kunz} {et~al.}(2011){Kunz}, {Schekochihin}, {Cowley}, {Binney}, \&
  {Sanders}}]{kunz11}
{Kunz}, M.~W., {Schekochihin}, A.~A., {Cowley}, S.~C., {Binney}, J.~J., \&
  {Sanders}, J.~S. 2011, \mnras, 410, 2446

\bibitem[{{Kunz} {et~al.}(2014){Kunz}, {Schekochihin}, \& {Stone}}]{kunz14}
{Kunz}, M.~W., {Schekochihin}, A.~A., \& {Stone}, J.~M. 2014, PhRvL, 112,
  205003

\bibitem[{{Laveder} {et~al.}(2011){Laveder}, {Marradi}, {Passot}, \&
  {Sulem}}]{laveder11}
{Laveder}, D., {Marradi}, L., {Passot}, T., \& {Sulem}, P.~L. 2011, GeoRL, 38,
  L17108

\bibitem[{{Marsch}(1986)}]{marsch86}
{Marsch}, E. 1986, \aap, 164, 77

\bibitem[{{Marsch} {et~al.}(1982){Marsch}, {Schwenn}, {Rosenbauer},
  {Muehlhaeuser}, {Pilipp}, \& {Neubauer}}]{marsch82}
{Marsch}, E., {Schwenn}, R., {Rosenbauer}, H., {et~al.} 1982, JGR, 87, 52

\bibitem[{{Marsch} \& {Tu}(1993)}]{marsch93}
{Marsch}, E., \& {Tu}, C.~Y. 1993, AnGeo, 11, 659

\bibitem[{{Marsch} \& {Verscharen}(2011)}]{marsch11}
{Marsch}, E., \& {Verscharen}, D. 2011, JPlPh, 77, 385

\bibitem[{{Marsch} {et~al.}(2006){Marsch}, {Zhao}, \& {Tu}}]{marsch06}
{Marsch}, E., {Zhao}, L., \& {Tu}, C.-Y. 2006, AnGeo, 24, 2057

\bibitem[{{Maruca} {et~al.}(2012){Maruca}, {Kasper}, \& {Gary}}]{maruca12}
{Maruca}, B.~A., {Kasper}, J.~C., \& {Gary}, S.~P. 2012, \apj, 748, 137

\bibitem[{{Matteini} {et~al.}(2012){Matteini}, {Hellinger}, {Landi},
  {Tr{\'a}vn{\'{\i}}{\v c}ek}, \& {Velli}}]{matteini12}
{Matteini}, L., {Hellinger}, P., {Landi}, S., {Tr{\'a}vn{\'{\i}}{\v c}ek},
  P.~M., \& {Velli}, M. 2012, SSRv, 172, 373

\bibitem[{{Matteini} {et~al.}(2007){Matteini}, {Landi}, {Hellinger},
  {Pantellini}, {Maksimovic}, {Velli}, {Goldstein}, \& {Marsch}}]{matteini07}
{Matteini}, L., {Landi}, S., {Hellinger}, P., {et~al.} 2007, GeoRL, 34, L20105

\bibitem[{{Montgomery} \& {Turner}(1981)}]{montgomery81}
{Montgomery}, D., \& {Turner}, L. 1981, PhFl, 24, 825

\bibitem[{{Narita} {et~al.}(2011){Narita}, {Glassmeier}, {Goldstein},
  {Motschmann}, \& {Sahraoui}}]{narita11}
{Narita}, Y., {Glassmeier}, K.-H., {Goldstein}, M.~L., {Motschmann}, U., \&
  {Sahraoui}, F. 2011, AnGeo, 29, 1731

\bibitem[{{Narita} \& {Marsch}(2015)}]{narita15}
{Narita}, Y., \& {Marsch}, E. 2015, \apj, 805, 24

\bibitem[{{Oughton} {et~al.}(1998){Oughton}, {Matthaeus}, \&
  {Ghosh}}]{oughton98}
{Oughton}, S., {Matthaeus}, W.~H., \& {Ghosh}, S. 1998, PhPl, 5, 4235

\bibitem[{{Oughton} {et~al.}(1994){Oughton}, {Priest}, \&
  {Matthaeus}}]{oughton94}
{Oughton}, S., {Priest}, E.~R., \& {Matthaeus}, W.~H. 1994, JFM, 280, 95

\bibitem[{{Quest} \& {Shapiro}(1996)}]{quest96}
{Quest}, K.~B., \& {Shapiro}, V.~D. 1996, JGR, 101, 24457

\bibitem[{{Riquelme} {et~al.}(2015){Riquelme}, {Quataert}, \&
  {Verscharen}}]{riquelme15}
{Riquelme}, M.~A., {Quataert}, E., \& {Verscharen}, D. 2015, \apj, 800, 27

\bibitem[{{Rosin} {et~al.}(2011){Rosin}, {Schekochihin}, {Rincon}, \&
  {Cowley}}]{rosin11}
{Rosin}, M.~S., {Schekochihin}, A.~A., {Rincon}, F., \& {Cowley}, S.~C. 2011,
  \mnras, 413, 7

\bibitem[{{Rudakov} \& {Sagdeev}(1961)}]{rudakov61}
{Rudakov}, L.~I., \& {Sagdeev}, R.~Z. 1961, SPhD, 6, 415

\bibitem[{{Sagdeev} \& {Shafranov}(1961)}]{sagdeev61}
{Sagdeev}, R.~Z., \& {Shafranov}, V.~D. 1961, Sov.~Phys.~JETP, 12, 130

\bibitem[{{Sahraoui} {et~al.}(2010){Sahraoui}, {Goldstein}, {Belmont}, {Canu},
  \& {Rezeau}}]{sahraoui10}
{Sahraoui}, F., {Goldstein}, M.~L., {Belmont}, G., {Canu}, P., \& {Rezeau}, L.
  2010, PhRvL, 105, 131101

\bibitem[{{Salem} {et~al.}(2012){Salem}, {Howes}, {Sundkvist}, {Bale},
  {Chaston}, {Chen}, \& {Mozer}}]{salem12}
{Salem}, C.~S., {Howes}, G.~G., {Sundkvist}, D., {et~al.} 2012, \apjl, 745, L9

\bibitem[{{Schekochihin} \& {Cowley}(2006)}]{schekochihin06}
{Schekochihin}, A.~A., \& {Cowley}, S.~C. 2006, PhPl, 13, 056501

\bibitem[{{Schekochihin} {et~al.}(2009){Schekochihin}, {Cowley}, {Dorland},
  {Hammett}, {Howes}, {Quataert}, \& {Tatsuno}}]{schekochihin09}
{Schekochihin}, A.~A., {Cowley}, S.~C., {Dorland}, W., {et~al.} 2009, \apjs,
  182, 310

\bibitem[{{Schekochihin} {et~al.}(2005){Schekochihin}, {Cowley}, {Kulsrud},
  {Hammett}, \& {Sharma}}]{schekochihin05}
{Schekochihin}, A.~A., {Cowley}, S.~C., {Kulsrud}, R.~M., {Hammett}, G.~W., \&
  {Sharma}, P. 2005, \apj, 629, 139

\bibitem[{{Schekochihin} {et~al.}(2016){Schekochihin}, {Parker}, {Highcock},
  {Dellar}, {Dorland}, \& {Hammett}}]{schekochihin16}
{Schekochihin}, A.~A., {Parker}, J.~T., {Highcock}, E.~G., {et~al.} 2016,
  JPlPh, 82, 905820212 (47 pages)

\bibitem[{{Servidio} {et~al.}(2014){Servidio}, {Osman}, {Valentini}, {Perrone},
  {Califano}, {Chapman}, {Matthaeus}, \& {Veltri}}]{servidio14}
{Servidio}, S., {Osman}, K.~T., {Valentini}, F., {et~al.} 2014, \apjl, 781, L27

\bibitem[{{Servidio} {et~al.}(2015){Servidio}, {Valentini}, {Perrone}, {Greco},
  {Califano}, {Matthaeus}, \& {Veltri}}]{servidio15}
{Servidio}, S., {Valentini}, F., {Perrone}, D., {et~al.} 2015, JPlPh, 81,
  325810107

\bibitem[{{Sharma} {et~al.}(2006){Sharma}, {Hammett}, {Quataert}, \&
  {Stone}}]{sharma06}
{Sharma}, P., {Hammett}, G.~W., {Quataert}, E., \& {Stone}, J.~M. 2006, \apj,
  637, 952

\bibitem[{{Southwood} \& {Kivelson}(1993)}]{southwood93}
{Southwood}, D.~J., \& {Kivelson}, M.~G. 1993, JGR, 98, 9181

\bibitem[{{Stix}(1992)}]{stix92}
{Stix}, T.~H. 1992, {Waves in plasmas, American Institute of Physics, New York,
  NY, USA}

\bibitem[{{Tajiri}(1967)}]{tajiri67}
{Tajiri}, M. 1967, JPSJ, 22, 1482

\bibitem[{{Taylor}(1938)}]{taylor38}
{Taylor}, G.~I. 1938, Proceedings of the Royal Society of London Series A, 164,
  476

\bibitem[{{Tu} \& {Marsch}(1994)}]{tu94}
{Tu}, C.-Y., \& {Marsch}, E. 1994, JGR, 99, 21

\bibitem[{{Tu} \& {Marsch}(1995)}]{tu95}
---. 1995, SSRv, 73, 1

\bibitem[{{Vellante} \& {Lazarus}(1987)}]{vellante87}
{Vellante}, M., \& {Lazarus}, A.~J. 1987, JGR, 92, 9893

\bibitem[{{Verscharen} {et~al.}(2013{\natexlab{a}}){Verscharen}, {Bourouaine},
  \& {Chandran}}]{verscharen13b}
{Verscharen}, D., {Bourouaine}, S., \& {Chandran}, B.~D.~G. 2013{\natexlab{a}},
  \apj, 773, 163

\bibitem[{{Verscharen} {et~al.}(2013{\natexlab{b}}){Verscharen}, {Bourouaine},
  {Chandran}, \& {Maruca}}]{verscharen13a}
{Verscharen}, D., {Bourouaine}, S., {Chandran}, B.~D.~G., \& {Maruca}, B.~A.
  2013{\natexlab{b}}, \apj, 773, 8

\bibitem[{{Verscharen} \& {Chandran}(2013)}]{verscharen13}
{Verscharen}, D., \& {Chandran}, B.~D.~G. 2013, \apj, 764, 88

\bibitem[{{Verscharen} \& {Marsch}(2011)}]{verscharen11}
{Verscharen}, D., \& {Marsch}, E. 2011, JPlPh, 77, 693

\bibitem[{{Verscharen} {et~al.}(2012){Verscharen}, {Marsch}, {Motschmann}, \&
  {M{\"u}ller}}]{verscharen12}
{Verscharen}, D., {Marsch}, E., {Motschmann}, U., \& {M{\"u}ller}, J. 2012,
  PhPl, 19, 022305

\bibitem[{{Wicks} {et~al.}(2011){Wicks}, {Horbury}, {Chen}, \&
  {Schekochihin}}]{wicks11}
{Wicks}, R.~T., {Horbury}, T.~S., {Chen}, C.~H.~K., \& {Schekochihin}, A.~A.
  2011, PhRvL, 106, 045001

\bibitem[{{Yao} {et~al.}(2011){Yao}, {He}, {Marsch}, {Tu}, {Pedersen},
  {R{\`e}me}, \& {Trotignon}}]{yao11}
{Yao}, S., {He}, J.-S., {Marsch}, E., {et~al.} 2011, \apj, 728, 146

\bibitem[{{Yao} {et~al.}(2013{\natexlab{a}}){Yao}, {He}, {Tu}, {Wang}, \&
  {Marsch}}]{yao13}
{Yao}, S., {He}, J.-S., {Tu}, C.-Y., {Wang}, L.-H., \& {Marsch}, E.
  2013{\natexlab{a}}, \apj, 776, 94

\bibitem[{{Yao} {et~al.}(2013{\natexlab{b}}){Yao}, {He}, {Tu}, {Wang}, \&
  {Marsch}}]{yao13a}
---. 2013{\natexlab{b}}, \apj, 774, 59

\bibitem[{{Yoon} \& {Seough}(2014)}]{yoon14}
{Yoon}, P.~H., \& {Seough}, J. 2014, JGR, 119, 7108

\bibitem[{{Zank} {et~al.}(1990){Zank}, {Matthaeus}, \& {Klein}}]{zank90}
{Zank}, G.~P., {Matthaeus}, W.~H., \& {Klein}, L.~W. 1990, GeoRL, 17, 1239

\end{thebibliography}

\end{document}